\documentclass{article}

\usepackage{PRIMEarxiv}

\usepackage[utf8]{inputenc} 
\usepackage[T1]{fontenc}    
\usepackage{hyperref}       
\usepackage{url}            
\usepackage{booktabs}       
\usepackage{amsfonts}       
\usepackage{nicefrac}       
\usepackage{microtype}      
\usepackage{lipsum}
\usepackage{fancyhdr}       
\usepackage{graphicx}       
\graphicspath{{media/}}     

\usepackage{multirow}
\usepackage{CJKutf8}
\newcommand{\name}[1]{EasyLAN}

\title{A Human-Computer Collaborative Tool for Training a Single Large Language Model Agent into a Network through Few Examples}

\author{
  Lihang Pan\thanks{Both authors contributed equally to the paper.}\\
  Tsinghua University \\
  Beijing\\
  \texttt{plh18@mails.tsinghua.edu.cn} \\
   \And
  Yuxuan Li\footnotemark[1]\\
  Tsinghua University \\
  Beijing\\
  \texttt{yuxuan-l20@mails.tsinghua.edu.cn} \\
  \And
  Chun Yu\thanks{Indicates the corresponding author.}\\
  Tsinghua University \\
  Beijing\\
  \texttt{chunyu@mail.tsinghua.edu.cn} \\
  \And
  Yuanchun Shi\\
  Tsinghua University \\
  Beijing\\
  \texttt{shiyc@tsinghua.edu.cn} \\
}

\begin{document}
\maketitle

\begin{abstract}
The capabilities of a single large language model (LLM) agent for solving a complex task are limited. Connecting multiple LLM agents to a network can effectively improve overall performance. However, building an LLM agent network (LAN) requires a substantial amount of time and effort. In this paper, we introduce EasyLAN, a human-computer collaborative tool that helps developers construct LANs. EasyLAN initially generates a LAN containing only one agent based on the description of the desired task. Subsequently, EasyLAN leverages a few training examples to update the LAN. For each example, EasyLAN models the gap between the output and the ground truth and identifies the causes of the errors. These errors are addressed through carefully designed strategies. Users can intervene in EasyLAN's workflow or directly modify the LAN. Eventually, the LAN evolves from a single agent to a network of LLM agents. The experimental results indicate that developers can rapidly construct LANs with good performance.
\end{abstract}


\begin{figure}[ht]
  \includegraphics[width=\textwidth]{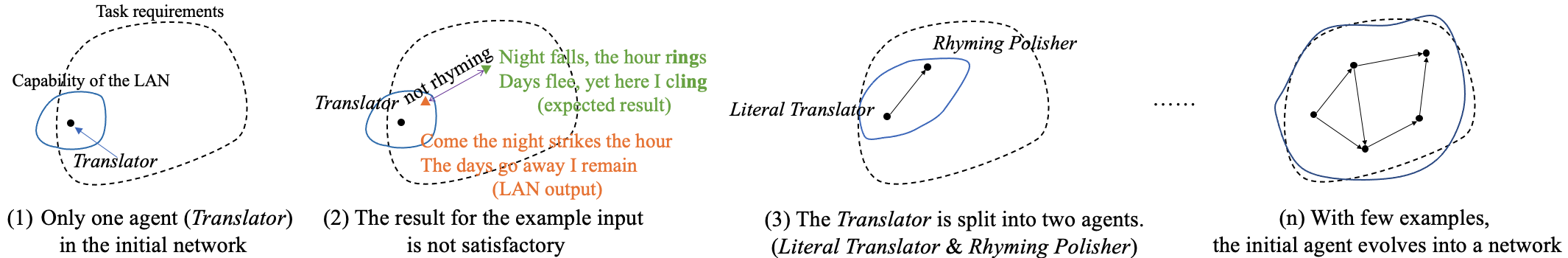}
  \caption{How \name{} trains a task-oriented LLM agent network (LAN) from a single LLM agent. (1) \name{} auto-generates an initial LAN that only contains a single LLM agent based on the task (e.g., translating French to English). A significant gap exists between the capabilities of the initial LAN and the task requirements. (2) A training example consists of an input and a ground truth. For a given training example, \name{} identifies discrepancies between the LAN's output and the expected output. For instance, when the input is a line of French poetry, "Vienne la nuit sonne l'heure, les jours s'en vont je demeure", the LAN fails to translate the text accurately while preserving the original rhyming scheme. (3) \name{} identifies the cause of the discrepancies and updates the LAN with respect to both the network architecture (e.g., splitting \textit{Translator} into \textit{Literal Translator} and \textit{Rhyming Polisher}) and agent contents (e.g., adjusting the functionality of an agent). (n) \name{} iterates over a small set of training examples and constructs a satisfactory LAN.}  \label{fig:teaser}
\end{figure}

\section{Introduction}
The capabilities of Large Language Models (LLMs) in handling complex tasks are limited. A common solution is to decompose a complex task into sub-tasks \cite{khot2022decomposed, zhou2022least, dua-etal-2022-successive}, each managed by an LLM-driven agent. These agents connect and form a network, collaboratively accomplishing the complex task. However, existing LLM agent networks (LANs) require manual editing by developers \cite{wu2022promptchainer}, demanding significant time and effort in the network's design, testing, and modification. Developers must (1) estimate LLM capabilities to determine the task division and interconnections among agents \cite{wu2022ai} and (2) inspect each agent to identify and resolve issues that lead to suboptimal performance. Human-computer collaboration \cite{kochhar1990user, kochhar1994interaction} is a promising solution that leverages artificial intelligence (AI) to alleviate physical and mental burdens. Nevertheless, this collaborative paradigm has not been applied in the design and development of LANs.

In this paper, we introduce \name{}, a human-computer collaborative editing tool for developing task-oriented LLM agent networks (LANs). The most significant feature is the "few-example-driven" paradigm for LAN construction. \name{} alleviates the need for proactive decomposition of the complex task during LAN design and reduces the effort for reactive inspection and modification of agents during LAN debugging. As illustrated in Figure \ref{fig:teaser}, \name{} initiates the LAN with a single agent based on a brief description of the complex task. For each training example, \name{} compares the actual and expected output to identify the limitations and the root causes within the LAN. \name{} then formulates and implements updates, improving the network's capabilities to accommodate the given example. The human-computer collaboration is manifested in the following way: the LAN developer supervises how \name{} executes the workflow mentioned above. When anomalies or errors occur, they can intervene and correct the system's actions to ensure appropriate execution.

To implement \name{}, we designed the internal structure of agents, which consists of input, control, execution, and output modules. The connection of input and output modules between different agents forms the network. The control module evaluates and decides whether the agent should be activated, while the execution module computes the agent output. Both of them leverage LLMs for computation and contain updatable knowledge bases and example repositories for few-shot learning. We designed strategies for updating the LAN and a pipeline for identifying error causes and selecting corresponding strategies. A user interface has been implemented in the browser to facilitate user inspection and control over the update process. The user can directly edit the LAN through GUI interactions (e.g., drag and drop) when necessary. We conducted a user study (N=12) to evaluate the usability of \name{}. The experimental results showed that \name{} can help users reduce interaction time by 39.3\% while improving the performance of the constructed LANs by 39.8\%.

In the remaining parts of this paper, we first review work related to our work. We then outline the design of the agent network (Section \ref{section: LAN design}), which serves as the output of \name{}. Subsequently, we describe how the user and \name{} collaboratively train the LAN, detailing: 1) the automated mechanisms which \name{} employs to update the LAN (Section \ref{section: automatic LAN update}); and 2) how the user inspects and intervenes in the update process (Section \ref{section: user interaction with EasyLAN}). Technical details of the system implementation are elaborated upon in Section \ref{section: impl}, focusing on the design of prompts, as the effectiveness of \name{} and the LAN heavily relies on LLMs. Finally, we validate the performance of \name{} through an evaluation study.

\section{Related Work}
\subsection{Decomposing Complex Tasks and Connecting Multiple LLM Agents}
Large Language Models (LLMs) have been applied to a wide array of tasks such as code generation \cite{liu2023wants}, storytelling \cite{chung2022talebrush}, and command understanding \cite{kim2022stylette}. However, their performance is limited in real-world tasks that are inherently complex and require domain-specific knowledge \cite{dai2021knowledge, meng2022locating} or multi-step reasoning \cite{huang2022towards, qiao2022reasoning}. One promising approach to address these limitations is to explicitly employ multiple LLM-driven agents to collaborate on the tasks, simulating the human process of decomposing complex tasks and addressing them separately \cite{teevan2016productivity, bernstein2010soylent, zhang2021learning}. For instance, AI Chains \cite{wu2022ai} accomplish tasks like peer-review writing and personalized flashcard creation by chaining LLM prompts. In addition to performance gains, this approach has been proved to offer improved transparency and controllability \cite{wu2022ai} compared to embedding multi-step reasoning within a single LLM calculation (e.g., chain of thought \cite{wei2022chain}) or repetitive calculations with similar prompts (e.g., SayCan \cite{brohan2023can} and AutoCoT \cite{zhang2022automatic}).

There are two primary challenges in creating a network of LLM agents. First, the design of the network architecture, specifically the decomposition of complex tasks into subtasks, is not straightforward \cite{kim2017mechanical, teevan2016productivity, wu2022ai}. Users must consider global constraints \cite{zhang2012human, kaur2018creating} and break down tasks into actionable subtasks \cite{zacks2001event}. These subtasks then need to be allocated to appropriate entities (e.g., self-sourcing \cite{teevan2014selfsourcing}, friend-sourcing \cite{agapie2016plansourcing, morris2010people, rzeszotarski2014estimating}, and crowd-sourcing \cite{anjum2021crowdmot}) based on their requirements such as expertise and task familiarity \cite{teevan2016productivity, he2023interaction}. The difficulty is further amplified when constructing an LLM network, as users struggle to predict LLM capabilities in solving specific problems \cite{dang2022prompt}. Second, using existing graphical user interface (GUI) editing tools introduces additional interaction overhead. Although researchers have proposed GUI tools for editing LAN (e.g., PromptChainer \cite{wu2022promptchainer}), users are confined to constructing, testing, and modifying the network through tedious GUI operations. Existing tools do not allow for human-machine collaboration to reduce users' workload.

\name{} addresses the aforementioned challenges in two ways. On the one hand, it obviates the need for users to design the network, thus reducing users' cognitive load. \name{} automatically tailors both the network structure and the content of each agent according to the provided training samples. On the other hand, \name{} automates the process of network update. Even if \name{} makes an error, users can intervene in the update pipeline and correct its behaviors through minimal interactions, thereby decreasing their physical workload.

\subsection{Using Examples in LLM Applications}
Existing research leverages few-shot learning \cite{brown2020language, reynolds2021prompt} to introduce examples (pairs of input and output) in the prompts. These examples guide the LLM to generate more accurate results. This method has substantially lowered the barriers to prototype AI applications, making it particularly beneficial for those without AI expertise. Consequently, this approach has been broadly adopted in various fields, including translation \cite{reynolds2021prompt}, chatbots \cite{wang2023enabling}, end-user programming \cite{fiannaca2023programming}, and robot interaction \cite{murali2023improving}. Although some studies have optimized the construction \cite{lampinen2022can, su2022selective, deng2022rlprompt, zhang2022active}, selection \cite{agrawal2022context, liu2021makes, rubin2021learning}, and order \cite{lu2021fantastically, zhao2021calibrate} of these examples to enhance model performance \cite{dong2022survey}, they have not fundamentally changed how examples are employed within LLM-based applications.

One concern among researchers is what LLMs are capable of learning through few-shot learning \cite{min2022rethinking, garg2022can, webson2021prompt, o2021context}, a question critical to evaluating the method's reliability for real-world applications \cite{ye2022unreliability, shi2023large, turpin2023language}. While preliminary insights and conclusions exist for AI experts ( e.g., few-shot examples may provide input distribution and output space \cite{min2022rethinking}), there has been no comprehensive answer suitable for non-AI experts (e.g., human-computer interaction (HCI) researchers without an AI background). This gap often forces these users into a 'trial-and-error' approach when creating prompts \cite{dang2022prompt, jiang2022promptmaker}, reducing efficiency, result quality, and user controllability.

Unlike the aforementioned approaches, \name{} goes beyond just hardcoding the examples in the prompts. \name{} explicitly adjusts the network architecture and the knowledge within each agent based on the provided examples. \name{} enables a more comprehensive utilization of the example data, maximizing the information extracted from the examples. It also gives developers a more intuitive understanding of what their systems have learned from the examples and simplifies debugging.

\subsection{Human-Computer Collaborative Tools}
In traditional human-computer collaboration models, the computer typically offers optional hints or suggestions, aiding the human user in task completion. In such systems, the human user shoulders the workload predominantly. These types of tools have been widely utilized across various domains, including but not limited to, patient note-taking \cite{wilcox2009activenotes, zhou2017intelligent, wang2021brilliant}, storytelling \cite{sullivan2018tarot, zhang2022storybuddy}, and other creative endeavors \cite{davis2016empirically, wang2010idea, guzdial2019friend, kim2022fitvid, kim2015motif}. 

With the advancement of AI, an increasing number of tools replace human involvement entirely. An illustrative example is AI-assisted painting, which can autonomously generate a complete image from a user-provided text description or partially complete image \cite{fan2019collabdraw, gatys2017controlling, ha2017neural, karimi2019deep, lin2020your, oh2018lead}. This autonomy, however, often comes at the cost of user agency in the creative process. Users frequently find themselves in a cycle of iteratively adjusting inputs \cite{liu2023wants, wang2023reprompt} or expending additional effort in interactive modifications to obtain desired outputs \cite{pan2023human, elgohary2021nl}. This highlights a critical gap in balancing automation and user control in current AI systems.

Human-in-the-loop (HITL) \cite{stiennon2020learning, ziegler2019fine} can be considered a special kind of human-computer collaboration. Instead of the computers assisting humans, HITL emphasizes humans helping computing systems improve their capabilities with additional effort (e.g., providing annotations). However, in existing HITL systems, users cannot engage in the internal updating processes of the system. This results in a lack of interactivity \cite{mosqueira2023human} in HITL and makes it challenging to support high-level knowledge \cite{wu2022survey}.

In this paper, we adopt a human-computer collaboration tool that diverges from the approaches above: the computer autonomously executes tasks (e.g., updating the LAN) while the human supervises its execution and intervenes if necessary. This approach not only alleviates the user's workload but also enhances the transparency and controllability of the computing system \cite{wu2022ai}.

\section{LLM Agent Network (LAN) Design}\label{section: LAN design}
\begin{figure}[ht]
  \centering
  \includegraphics[width=\textwidth]{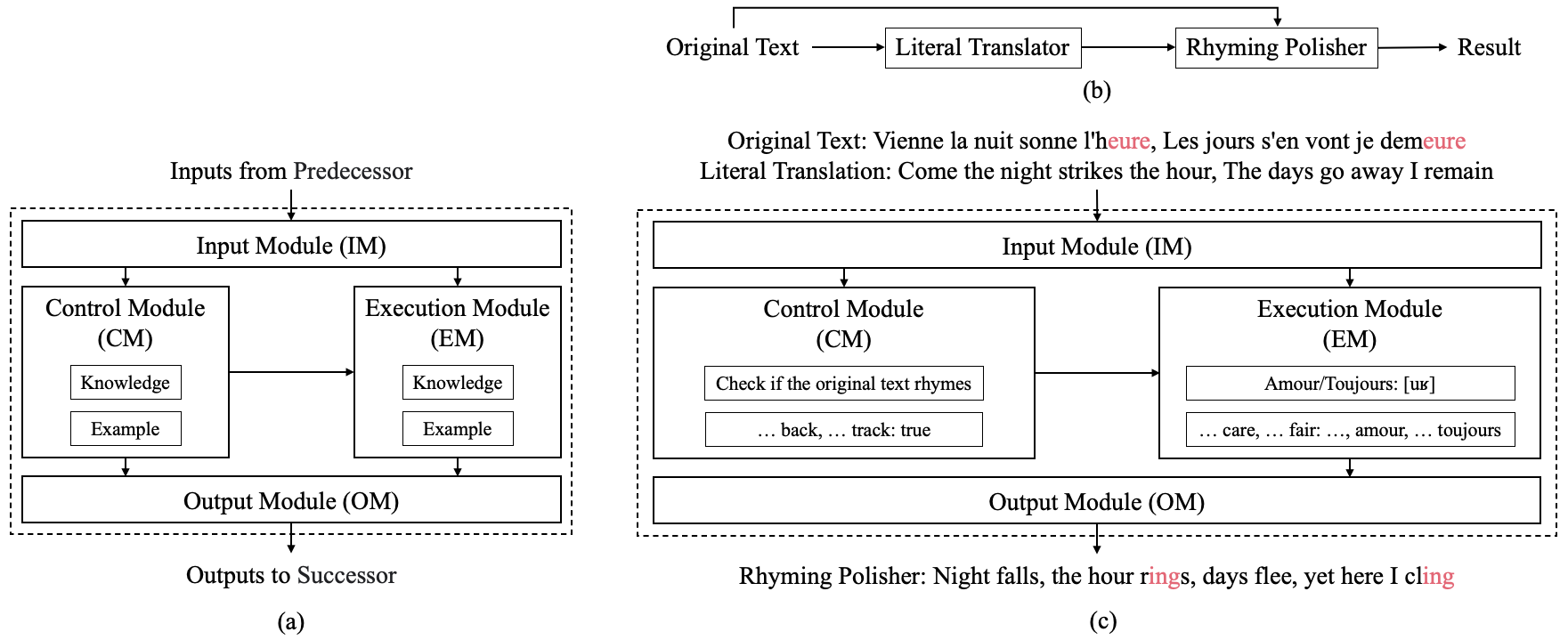}
  \caption{An overview of an agent in a LAN. (a) The modules inside an agent. (b) an example LAN. "Literal Translator" and "Rhyming Polisher" are agents in the LAN. (c) Details of the "Rhyming Polisher" in (b). It receives the original French text and the output from the "Literal Translator" and computes a rhyming translation result."}
  \label{fig: agent overview}
\end{figure}
A LAN is composed of multiple interconnected agents. The agents are responsible for sub-tasks and collaboratively accomplish a specific complex task. As illustrated in Figure \ref{fig: agent overview}(a), an agent comprises four components: an input module, a control module, an execution module, and an output module. The input module receives data from the output modules of predecessor agents, thereby connecting the agents into a directed acyclic graph (DAG). The control and execution modules constitute the core of an agent. They leverage LLMs to accomplish their functionalities and are subject to optimization during LAN updates.

\subsection{Input Module (IM) \& Output Module (OM)}\label{section: IOM}
The input module (IM) receives results from the predecessor agents' output modules (OMs). It also accepts external inputs (e.g., the French text awaiting translation in Figure \ref{fig: agent overview}(c)). The OM produces varying outputs based on the agent's activation status\footnote{The activation status will be further elaborated in Section \ref{section: CM}.}:
\begin{enumerate}
    \item When the agent is not activated (i.e., the control module returns 'False'), the OM forwards the received inputs as its outputs.
    \item When the agent is activated (i.e., the control module returns 'True'), the OM outputs the results generated by the execution module.
\end{enumerate}

\subsection{Control Module (CM)}\label{section: CM}
The Control Module (CM) serves as a decision-maker, akin to a "router", to evaluate whether the current agent should be activated. It facilitates conditional activation of agents, allowing the LAN to adapt its flow in real time by selecting a directed trail within the network. For instance, as illustrated in Figure \ref{fig: agent overview}(c), the 'Rhyming Polisher' is activated when the CM identifies rhyming elements in the French text input ("l'heure" and "demeure"). Conversely, for non-rhyming inputs, the CM would not activate the agent. By embedding such decision-making logic within the agent itself, instead of adding an additional agent to the network, we substantially simplify the LAN's architecture. The CM's decision impacts the final output of an agent, as discussed in Section \ref{section: IOM}.

\begin{table}[ht]
\caption{CM Properties and their explanations}
\label{table: CM property}
\resizebox{\textwidth}{!}{%
\begin{tabular}{|c|l|}
\hline
Property &
  \multicolumn{1}{c|}{Explanation} \\ \hline
Enabled &
  \begin{tabular}[c]{@{}l@{}}Boolean. If this property is set to false, it indicates that the CM will activate the agent under any input. \\ This attribute is employed for critical agents within the LAN (e.g., Literal Translator) to reduce computational cost.\end{tabular} \\ \hline
\begin{tabular}[c]{@{}c@{}}Required\\ Predecessors\end{tabular} &
  A list of predecessor agents. If any of these agents are not activated, the CM will not activate its agent. \\ \hline
Knowledge &
  \begin{tabular}[c]{@{}l@{}}A knowledge base. The CM utilizes the knowledge to determine whether to activate its agent.\\ For example, "If the original text exhibits rhyming, the current agent should be activated." (Rhyming Polisher in Figure \ref{fig: agent overview}(c))\end{tabular} \\ \hline
Examples &
  \begin{tabular}[c]{@{}l@{}}An example list. The CM utilizes the examples to determine whether to activate its agent.\\ An example is composed of the input of the agent and the CM's result. For example:\\ Input: (from LAN) Bras levés raides pour blâmer, Dans faux gestes d'aimer; \\\:\:\:\:\:\:\:\:\:\:\:\:(from \textit{Literal Translator}) Arms raised stiff to blame, In false gestures of loving. \\ Result: true\end{tabular} \\ \hline
\end{tabular}%
}
\end{table}

Table \ref{table: CM property} outlines the properties of a CM. When enabled and provided that all required predecessor agents are activated, the CM leverages LLMs to evaluate whether the agent should be activated. This assessment is based on the external input, outputs from upstream agents, and the specific subtask the agent is designed to undertake. For details of the prompt used in the CM, please refer to Section \ref{section: LAN exec prompt}.

\subsection{Execution Module (EM)}
The agent completes its subtask using the execution module (EM). Table \ref{table: EM property} presents the properties of an EM. The EM utilizes LLMs to compute the agent's output based on the external input, outputs from upstream agents, and its designated task. For details of the prompt used in the CM, please refer to Section \ref{section: LAN exec prompt}.

\begin{table}[ht]
\caption{EM Properties and their explanations. All the examples come from the EM of the "Literal Translator" in Figure \ref{fig: agent overview}}
\label{table: EM property}
\resizebox{\textwidth}{!}{%
\begin{tabular}{|c|l|}
\hline
Property &
  \multicolumn{1}{c|}{Explanation} \\ \hline
\begin{tabular}[c]{@{}c@{}}Subtask\\ Description\end{tabular} &
  A string, describing the subtask the agent undertakes. For example:  "Translating its literal meaning from the input into English" \\ \hline
\begin{tabular}[c]{@{}c@{}}Output\\ Description\end{tabular} &
  A string, describing the output of the agent. For example: "A string, indicating the literal translation" \\ \hline
Knowledge &
  A knowledge base. The EM utilizes the knowledge to execute its subtask. For example, "levés means lifted in English" \\ \hline
Examples &
  \begin{tabular}[c]{@{}l@{}}An example list. The EM utilizes the examples to decide how to execute its subtask.\\ An example is composed of the external output and the EM's result. For example:\\ Input: (from LAN) Bras levés raides pour blâmer, Dans faux gestes d'aimer; \\\:\:\:\:\:\:\:\:\:\:\:\:(from \textit{Literal Translator}) Arms raised stiff to blame, In false gestures of loving. \\  Result: Arms raised stiff to blame, In false gestures of loving\end{tabular} \\ \hline
\end{tabular}%
}
\end{table}

\section{System Design}
An LLM agent network (LAN) decomposes a complex task into multiple sub-tasks, each of which is managed by an individual agent. \name{} facilitates the user in developing a LAN for a complex task. The user is required to provide only 1) a natural language description of the desired task and 2) a small set of training examples comprising both inputs and expected outputs. \name{} employs an iterative approach to update the LAN. Throughout this process, users supervise the automated behaviors executed by \name{} and intervene as necessary. This section is organized as follows: first, we elucidate the automated pipeline responsible for LAN updates (Section \ref{section: automatic LAN update}), followed by an introduction on how the user can exert influence over \name{}'s workflow (Section \ref{section: user interaction with EasyLAN}).

\subsection{Automated Update of LAN}\label{section: automatic LAN update}
\subsubsection{Initialization of the LAN}
The initial agent network is comprised solely of a single agent. Its functionality is delineated by the user-provided task description, and its output is directly the outcome of that specified task. However, the performance of a single agent is inherently circumscribed due to its limitation of generating only end-to-end output.

\subsubsection{Pipeline to update a LAN}\label{section: update pipeline}
\begin{figure}[ht]
  \centering
  \includegraphics[width=\textwidth]{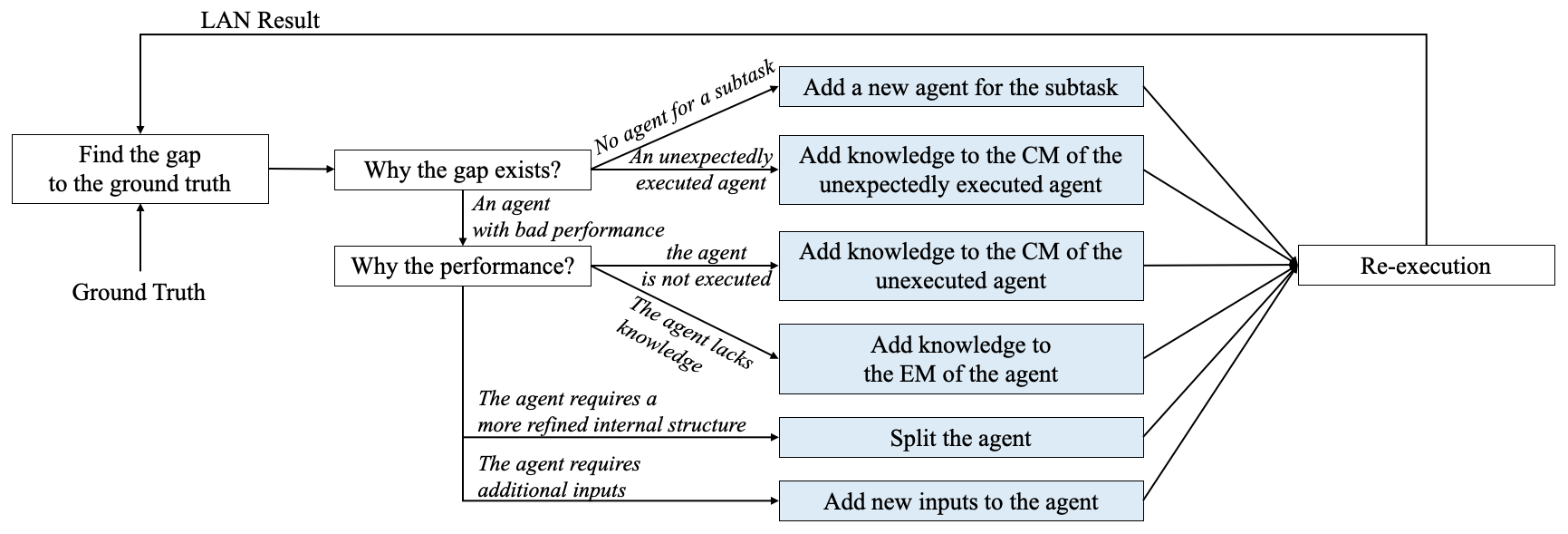}
  \caption{The pipeline for updating the LAN. The light blue rectangles indicate update strategies.}
  \label{fig: pipeline}
\end{figure}

Figure \ref{fig: pipeline} outlines the pipeline for updating the LAN. The pipeline comprises multiple iterations until the LAN can proficiently handle the current training input. At this time, \name{} adds the inputs and outputs of each agent's CM and EM to the corresponding module's example library, further ensuring the performance of the LAN.

During each iteration, \name{} selects an update strategy from Table \ref{table: update strategies} following the four steps below:
\begin{enumerate}
    \item (Step 1) \name{} calculates the gap between the LAN output and the expected ground truth. If the LAN fails to yield satisfactory results, it indicates that at least one constituent sub-task has been inadequately executed. \name{} identifies the most significant deficiency in the LAN and isolates the crucial sub-task.
    \item (Step 2) \name{} analyzes the reasons for the gap's existence, which fall into three categories:
    \begin{itemize}
        \item No agent is responsible for the sub-task.\name{} selects the strategy that creates a new agent and proceeds to Step 4.
        \item The sub-task should not be executed, but the corresponding agent is activated. \name{} chooses the strategy that updates the agent's CM knowledge to deactivate it and then moves to Step 4.
        \item An agent already manages the sub-task, but its performance is poor. \name{} proceeds to Step 3 to identify a more specific reason.
    \end{itemize}
    \item (Step 3) \name{} investigates the reasons behind the agent's poor performance and advances to Step 4. These reasons can be classified into four possibilities:
    \begin{itemize}
        \item The agent is not activated. \name{} selects the strategy to update the agent's CM knowledge to activate it.
        \item The agent lacks knowledge. \name{} chooses the strategy to update the agent's EM knowledge to enhance its performance.
        \item The agent requires a more refined internal structure. \name{} selects the strategy to split the agent into two or more agents.
        \item The agent needs additional inputs from other agents. \name{} opts for the strategy that adds new edges to the network, allowing the agent to receive the necessary inputs.
    \end{itemize}
    \item (Step 4) \name{} calculates the parameters for the chosen strategy. The LAN undergoes an update with this strategy, is re-executed to obtain a new result, and initiates a new iteration.
\end{enumerate}

\begin{table}[ht]
\centering
\caption{Update strategies and the underlying causes of error to rectify}
\label{table: update strategies}
\begin{tabular}{|c|l|}
\hline
Update Strategy &
  \multicolumn{1}{c|}{Cause of Error} \\ \hline
Add an agent &
  No agent is responsible for a sub-task. \\ \hline
Split an agent &
  \begin{tabular}[c]{@{}l@{}}A task is already managed by an agent,\\ but requires multiple steps or varying conditions to achieve a satisfactory output.\end{tabular} \\ \hline
\multirow{2}{*}{\begin{tabular}[c]{@{}c@{}}Add Knowledge\\ to the CM\end{tabular}} &
  A task is already managed by an agent, but the agent is not activated. \\ \cline{2-2} 
 &
  \begin{tabular}[c]{@{}l@{}}A sub-task that should not be executed is carried out,\\ indicating that an agent is erroneously activated.\end{tabular} \\ \hline
\begin{tabular}[c]{@{}c@{}}Add Knowledge\\ to the EM\end{tabular} &
  \begin{tabular}[c]{@{}l@{}}A task is already managed by an agent,\\ but requires additional knowledge to yield satisfactory output.\end{tabular} \\ \hline
\begin{tabular}[c]{@{}c@{}}Add inputs\\ to an Agent\end{tabular} &
  \begin{tabular}[c]{@{}l@{}}A task is already managed by an agent,\\ but requires additional inputs from other agents to yield satisfactory output.\end{tabular} \\ \hline
\end{tabular}%
\end{table}

\begin{figure}[ht]
  \centering
  \includegraphics[width=\textwidth]{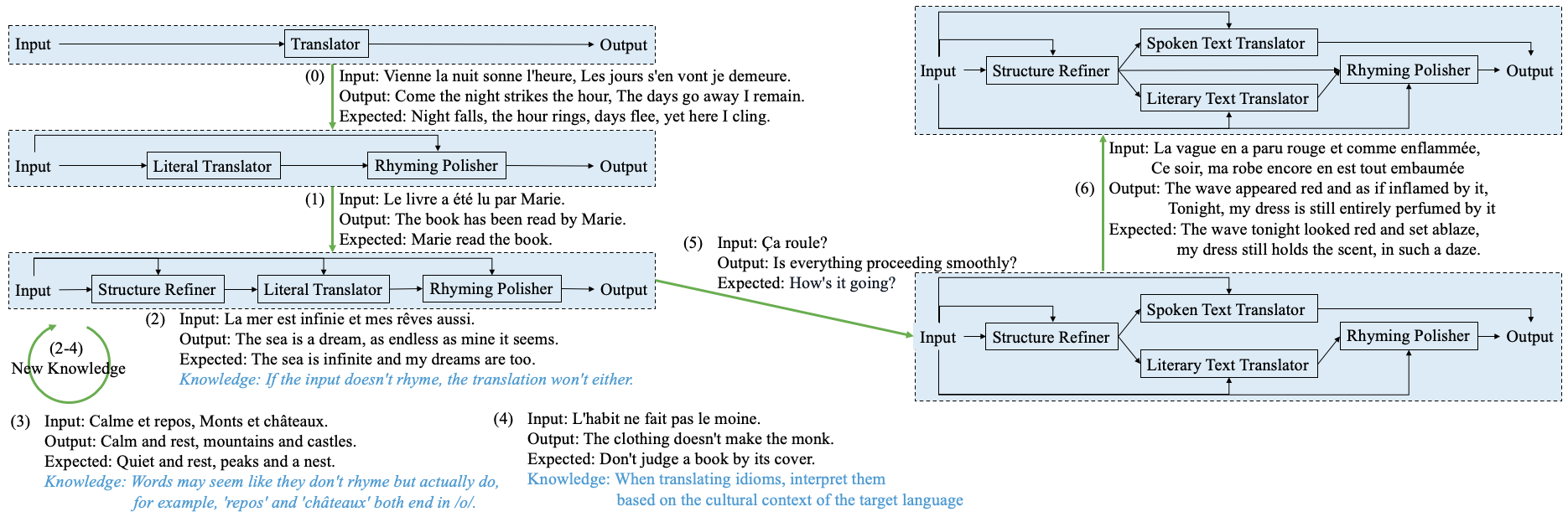}
  \caption{An example of how \name{} updates a LAN. (0) \name{} decomposes the \textit{Translator} into \textit{Literal Translator} and \textit{Rhyming Polisher} to ensure that the translation output can rhyme when necessary. (1) \name{} adds a \textit{Structure Refiner} to adjust the syntactic structure of sentences (e.g., converting passive voice to active voice). (2) \name{} adds knowledge to the CM of the \textit{Rhyming Polisher} to prevent unnecessary rhyming. (3) \name{} adds knowledge to the CM of the \textit{Rhyming Polisher} to better identify whether the input sentence is rhyming. (4) \name{} adds knowledge to the EM of the \textit{Literal Translator} to improve its capability to translate idioms. (5) \name{} splits the \textit{Literal Translator} into \textit{Spoken Text Translator} and \textit{Literary Text Translator} to better cater to diverse translation needs. (6) \name{} adds a connection from the \textit{Structure Refiner} to the \textit{Rhyming Polisher} to ensure that the output from the \textit{Rhyming Polisher} also adheres to the sentence structure defined by the \textit{Structure Refiner}.}
  \label{fig: pipeline example}
\end{figure}

\subsubsection{Details of the update strategies}
This section focuses on the specifics of various error causes and their respective update strategies.

\textbf{No agent is responsible for a sub-task}. \name{} creates a new agent for the sub-task and positions the agent appropriately within the network. As depicted in Figure \ref{fig: pipeline example}(1), French employs passive voice, which may result in redundancy when translated word-for-word into English. Therefore, \name{} adds a \textit{Structure Refiner} into the LAN to explicitly assess the sentence structure of translated results (e.g., active or passive voice). 

\textbf{A sub-task (handled by an agent) that should not be executed is carried out}. \name{} updates the knowledge of the CM to deactivate the mistakenly executed agent. As depicted in Figure \ref{fig: pipeline example}(2), the input sentence lacks rhyme, yet the \textit{Rhyming Polisher} is triggered and converts the output from the \textit{Literal Translator} into rhyming text. Rhyming carries side effects that may lead to semantic changes. Therefore, \name{} augments the knowledge of the \textit{Rhyming Polisher}'s CM to explicitly require that the \textit{Rhyming Polisher} should not execute if the original text does not rhyme.

\textbf{A sub-task is already managed by an agent, but its performance is suboptimal}. \name{} assesses the reasons behind the bad performance and selects corresponding strategies:
\begin{enumerate}
    \item If the agent is not activated, \name{} updates the knowledge of its CM to activate it. As depicted in Figure \ref{fig: pipeline example}(3), the LLM erroneously perceives "repos" and "châteaux" as ending with different vowels and deactivates the \textit{Rhyming Polisher}. \name{} thus adds knowledge to the CM to enforce its activation.
    \item If the agent requires additional knowledge to yield satisfactory results, \name{} summarizes and adds the required knowledge to the EM of the agent. As illustrated in Figure \ref{fig: pipeline example}(4), the input text carries specific cultural meanings ("one must not trust appearances"), yet the \textit{Literal Translator}'s output lacks this nuance. Consequently, \name{} enriches the knowledge within the \textit{Literal Translator}'s EM to enhance its performance.
    \item If the agent requires a more intricate structure, \name{} splits the agent into two or more separate agents and determines their interconnections. \name{} ensures that this division maintains the agent's semantic consistency. The knowledge of the original agent will be redistributed among the new agents. \name{} supports two kinds of division: (1) sequential division, which breaks down the sub-task into finer-grained steps (e.g., dividing the \textit{Translator} into the \textit{Literal Translate} and the \textit{Rhyme Optimization}, as shown in Figure \ref{fig: pipeline example}(0)); (2) parallel division, separates the agent based on distinct conditions. For example, in Figure \ref{fig: pipeline example}(5), the \textit{Literal Translator} is split into the \textit{Literary Text Translator} and the \textit{Spoken Text Translator} to handle different literary styles.
    \item If the agent requires additional inputs from other agents to yield satisfactory results, \name{} adds new edges to the network to ensure that the agent receives the necessary inputs. These additional inputs should already be computed within the agent network; otherwise, \name{} would create an agent responsible for their computation. As illustrated in Figure \ref{fig: pipeline example}(6), although the \textit{Structure Refiner} has already stipulated the use of an active voice in the translation result, this information was not conveyed to the \textit{Rhyming Polisher}, leading to it once again altering the result to a passive voice. To rectify this issue, \name{} establishes a connection from the \textit{Structure Refiner} to the \textit{Rhyming Polisher}.
\end{enumerate}

\subsubsection{Ensuring the accuracy of previous training inputs}
\name{}'s application of an update strategy should not compromise the accuracy of past training inputs. We adopt the following approaches to ensure this:
\begin{enumerate}
    \item \name{} does not encompass disruptive update strategies, such as agent deletion, connection removal, knowledge removal, or task modification. The current LAN is generally satisfactory and can correctly process previous training inputs. These strategies remove useful components from the network without providing any compensation.
    
    \item When the update strategy is adding knowledge (whether to the CM or the EM), \name{} employs few-shot learning to ensure the accuracy of historical training samples. As mentioned in Section \ref{section: update pipeline}, when the LAN can correctly handle a training sample, its operational state (e.g., the inputs and outputs of each agent's CM and EM) is recorded and added to the example repository. Therefore, \name{} does not need to take any additional actions.
    
    \item When the update strategy is adding inputs, the few-shot learning mechanism still works. This is because the old inputs remain present, allowing the LAN to fully utilize the examples based on them. Therefore, \name{} does not need to take any additional actions.

    \item When the update strategy is adding an agent, \name{} adds negative examples to the new agent's CM to ensure that the new agent will not be activated when processing historical training inputs. By taking this additional approach, we can ensure that the set of agents activated when handling historical inputs remains consistent, and therefore the final output also remains unchanged.
    
    \item When the update strategy is splitting an agent, \name{}'s additional actions are quite complex. For a given historical input, if the agent being split should not be activated, we simply add negative examples to the CMs of the newly created agents to ensure they will not be activated either. When the agent being split needs to be activated, \name{} also splits the examples of the original agent:
    \begin{enumerate}
        \item Splitting into multiple parallel agents. \name{} utilizes LLM to select an agent from the new agents. Then \name{} adds examples to the CMs of the new agents so that only the selected agent will be activated when processing the given history input. Finally, \name{} adds a new example to the EM of the selected agent to ensure that the output of that agent matches the output of the original agent.
        
        \item Splitting into multiple sequential agents. We only concern the output of the last agent. Therefore, \name{} executes this historical input in the new LAN: for non-last agents, \name{} adds the execution results as an example to the EMs; for the last agent, \name{} adds the execution result of the original agent to its EM.
    \end{enumerate}
\end{enumerate}

Note that, as outlined in Section \ref{section: user interaction with EasyLAN}, developers have the capability to manually edit the LAN. Given the potential complexity of these manual interventions, automatically ensuring the accuracy of previous training examples falls outside the scope of this paper. 

\subsection{User Interaction with \name{}}\label{section: user interaction with EasyLAN}

User interaction is divided into two categories: (1) Supervision of the automatic update pipeline by \name{}: If users believe that \name{} has made errors during this process, they can manually intervene. \name{} will continue executing unfinished steps following user intervention. (2) Manual editing of the LAN: Users can edit the LAN at any time. 

\begin{figure}[ht]
  \centering
  \includegraphics[width=\textwidth]{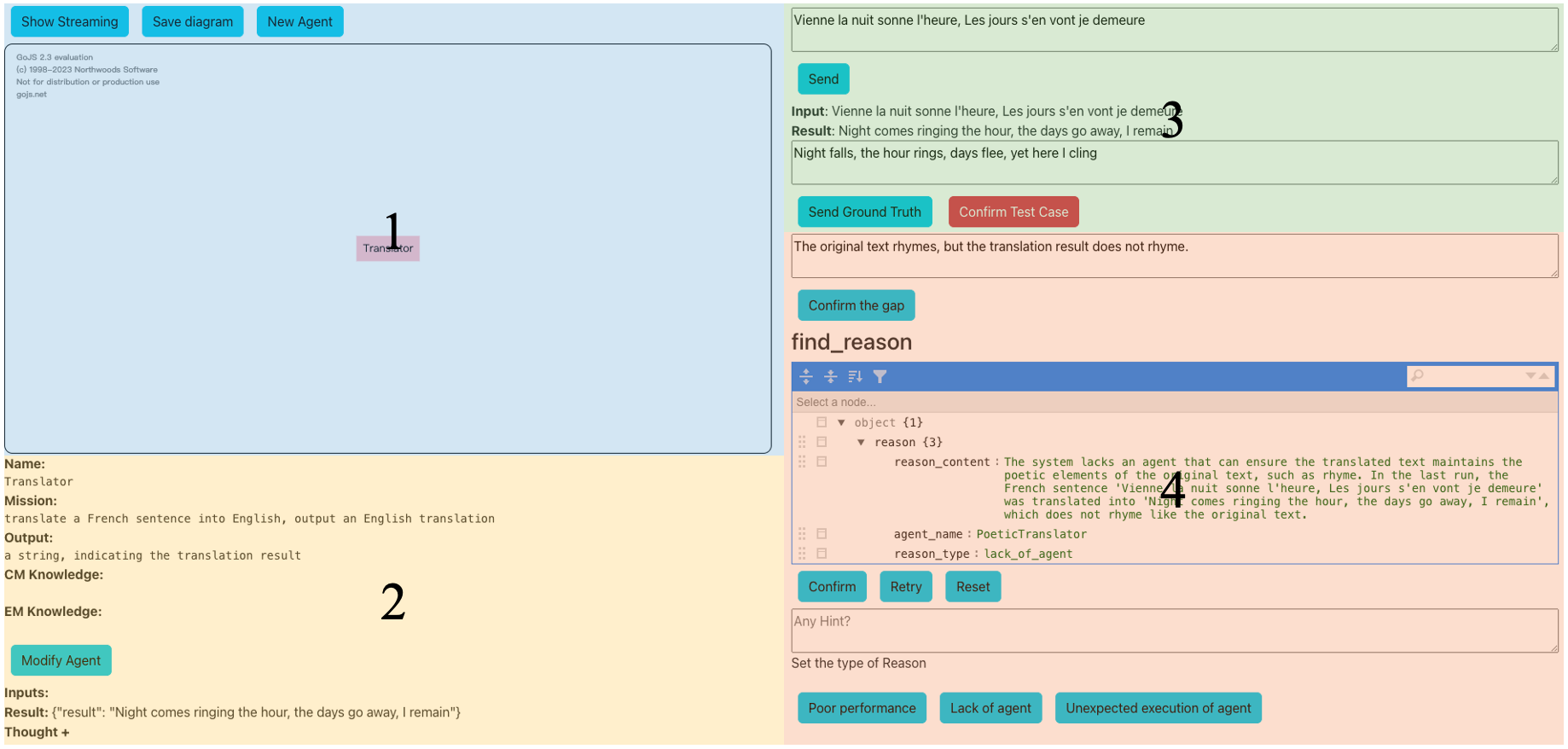}
  \caption{The user interface. Region 1 allows users to inspect and modify the LAN structure. Agents can be selected by clicking the pink rectangles. Region 2 facilitates inspection and editing of the selected agent's properties. In Region 3, users can provide training examples to \name{}. Region 4 offers insights into and intervention options for \name{}'s automated LAN update workflow.}
  \label{fig: ui}
\end{figure}

Figure \ref{fig: ui} displays the interaction interface, featuring Region 1 with a diagram providing an overview of the structure of the LAN. Users can select any agent in the diagram, with its detailed information appearing in Region 2. Region 3 allows users to provide training samples (both input and output) to the system, while Region 4 enables users to monitor and intervene in \name{}'s LAN updating process.

\subsubsection{Supervision and intervention in \name{}}
\begin{figure}[ht]
  \centering
  \includegraphics[width=0.75\textwidth]{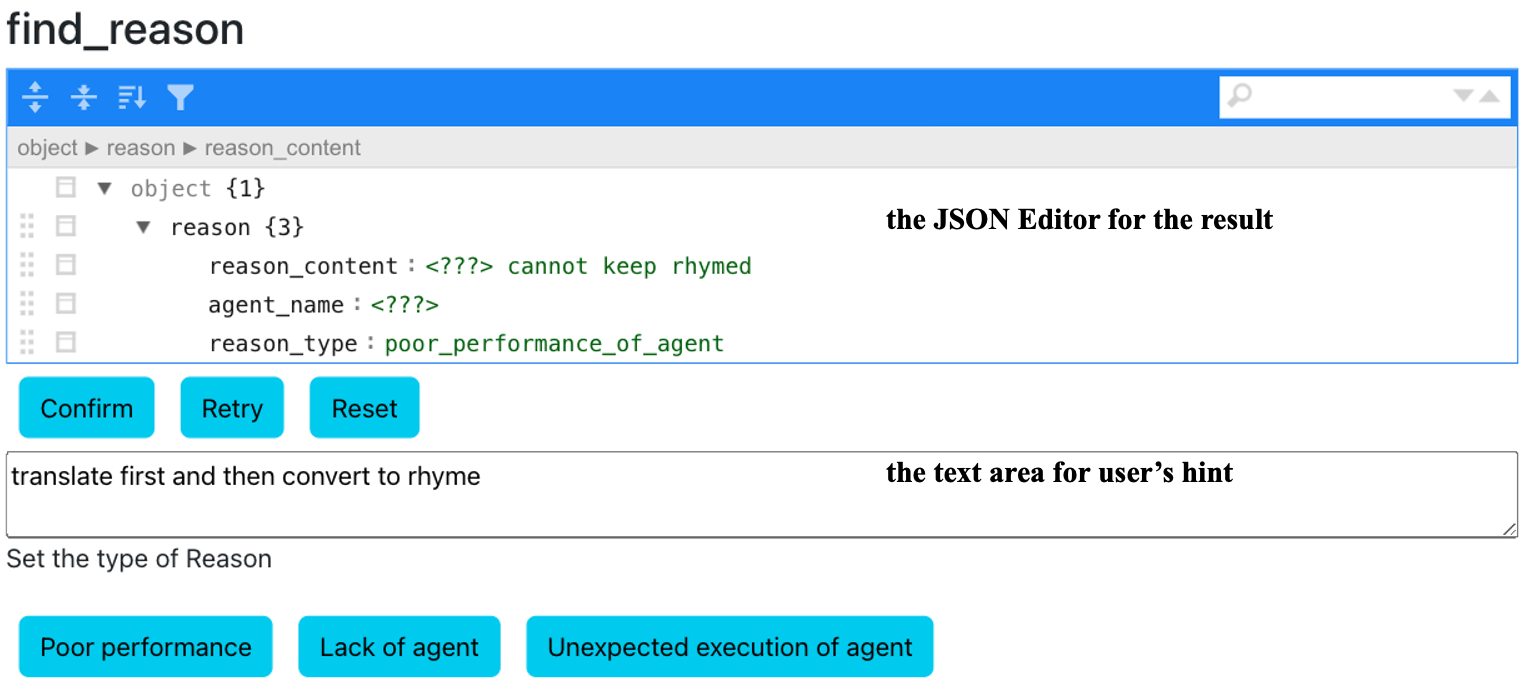}
  \caption{How the user intervenes in the pipeline of \name{} updating the LAN. The user can directly modify the result within the JSON editor and utilize placeholders (<???>) to allow \name{} to auto-complete them. They can also enter hints in the text area to guide \name{} in adjusting the results.}
  \label{fig: user interaction}
\end{figure}

\name{} displays results at each step of the LAN update process. As shown in Figure \ref{fig: ui}-4, the most significant discrepancy between the LAN's output and the ground truth is whether the translation result maintains rhyme. \name{} first calculates the underlying cause of this issue, which is the absence of an agent called \textit{PoeticTranslator}, and presents the result for user review. If the user believes the current step is correct, they can click "confirm" to proceed to the next step. In cases where the user finds \name{}'s results unsatisfactory, they can intervene in two ways (separately or simultaneously, as shown in Figure \ref{fig: user interaction}) and click "retry" to have the system recalculate based on their intervention.

\textbf{Way 1: manual modification of \name{}'s output}.As shown in Figure \ref{fig: ui}-4 \& \ref{fig: user interaction}, we utilize a JSON editor to present the result of the current step, allowing the user to edit them directly within the editor. The user can focus solely on important details and use placeholders (<???>) in non-essential properties, which \name{} will auto-complete. We provide keyboard shortcuts to facilitate this process, enabling users to (1) click a field (e.g., agent\_name in Figure \ref{fig: user interaction}) to set its value as a placeholder and (2) insert a placeholder where the cursor is. Additionally, we have implemented buttons to assist users in configuring the values of some key fields. For example, users can directly click "Poor performance" or "Lack of agents" to quickly set the value of the field "reason\_type," which indicates the cause of the error.

\textbf{Way 2: providing hints for \name{} to adjust its computation}. In this approach, the user does not tediously modify the details of \name{}'s results but can instead influence \name{}'s computation results through concise, high-level natural language descriptions. This method offers significant advantages, especially when \name{}'s computations are complex. For instance, when \name{} splits an agent, it calculates all the properties of the new agents (i.e., Table \ref{table: CM property} \& \ref{table: EM property}). If the user is dissatisfied with the result, directly modifying these properties, even a few, can be cumbersome. In contrast, the user can directly describe their expected splitting result here (e.g., "translate first and then convert to rhyme") and request \name{} to retry accordingly.

\subsubsection{Manual modification of the LAN}
The user can manually edit the agent network in Region 1. We support four operations: (1) Creating a new agent by clicking the "New agent" button. Detailed information about the new agent can be modified in Region 2. (2) Deleting an agent by selecting it and pressing the "Delete" key on the keyboard. This action also removes both the incoming and outgoing edges of the agent. (3) Connecting two agents by dragging the starting agent onto the target agent. (4) Deleting a connection between agents by selecting the edge and pressing the "Delete" key on the keyboard. When the user selects an agent in the diagram in Region 1, they can manually edit the agent's name, task description, output description, and knowledge of CM and EM in Region 2.

\name{} automatically verifies the editing results. The LAN cannot be saved if any of the following conditions are satisfied: (1) the network contains cycles, (2) any agent's name, task description, or output description is empty, or (3) two agents have duplicate names.

\section{Implementation}\label{section: impl}
\subsection{LAN execution}\label{section: LAN exec prompt}
\begin{figure}[ht]
  \centering
  \includegraphics[width=\textwidth]{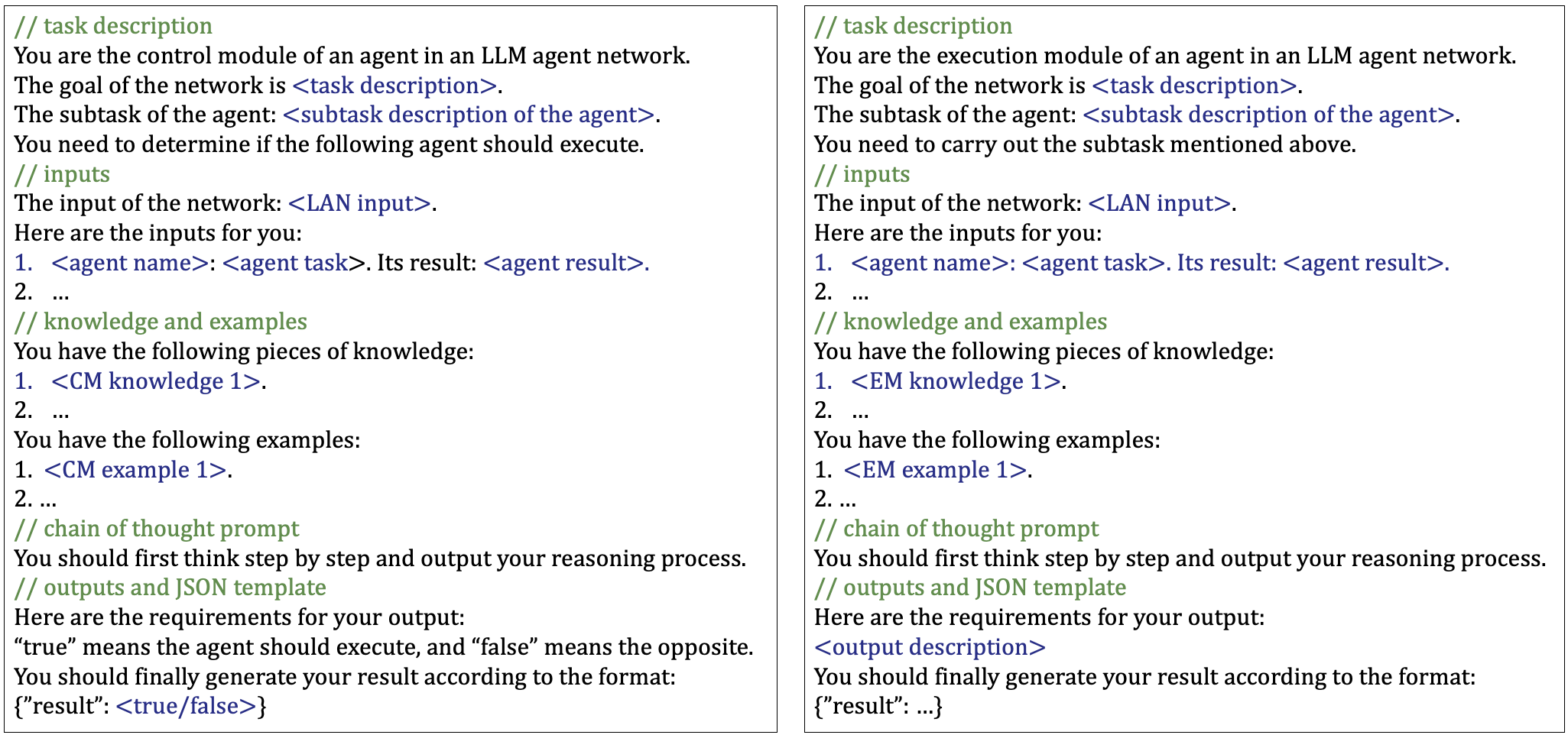}
  \caption{The prompts used in agent execution. Left: the prompt used by the CM to determine whether the agent should be activated. Right: the prompt used by the EM to calculate the output of the agent. Lines starting with a double slash ("//") are comments. Angle brackets (<>) in the prompts denote placeholders that should be replaced with actual values.}
  \label{fig: prompt agent exec}
\end{figure}

\name{} executes each agent sequentially following a topological order. The output of the last agent executed will serve as the LAN's final output. For each agent, \name{} utilizes LLMs to determine its activation status. If activated, it proceeds with result computation. Figure \ref{fig: prompt agent exec} illustrates the prompt's structure, which comprises: (1) a task description, specifying the LLM's current task; (2) agent inputs, encompassing the LAN's input, the sub-tasks of the agent's predecessors, and their corresponding outputs; (3) knowledge and examples (if available); (4) a zero-shot Chain of Thought prompt to extract the LLM's reasoning process; (5) a JSON template that defines the desired output format. It is important to note that the LLM may not consistently adhere to the format, in which case \name{} will call the LLM again to adjust the output to meet the specified format.

\subsection{LAN Update}
In this section, we focus on the implementation of each step in the LAN update process described in Section \ref{section: automatic LAN update}, especially the prompts \name{} uses.

\textbf{Description of the LAN}. We incorporate LAN descriptions into the prompts to provide the LLM with a comprehensive understanding of the LAN's content. This facilitates the LLM in better identifying its shortcomings. As depicted in Section \ref{section: LAN desc prompt} in the Appendix, these descriptions encompass the following elements: (1) Information regarding all agents, including their names and task descriptions; (2) Data flow between agents, specifying the data's content, its source, and its destination. This data flow originates from the LAN's last execution and should be optimized to yield satisfactory LAN outputs. (3) The LAN's input and output.

\textbf{Description of the agent}. We include agent descriptions in the prompts to provide the LLM with comprehensive insights into the agent's specifics. This aids the LLM in making better determinations regarding agent modifications. As depicted in Section \ref{section: Agent desc prompt} in the Appendix, agent descriptions encompass the following elements: (1) agent information, including the agent's name, task description, and \textbf{output description}; (2) \textbf{knowledge and examples within the agent's CM}; (3) \textbf{knowledge and examples within the agent's EM}; (4) agent's inputs, \textbf{thought processes (involving both CM and EM)}, and outputs from the previous execution. The content in bold was not included when describing the LAN.

\textbf{Update steps}. As shown in Table \ref{table: step prompt}, the prompts used in each update step consist of the following components: (1) the task to be accomplished in the current step, which aligns with the discussion in Section \ref{section: automatic LAN update}; (2) the outputs of previous steps; (3) a description of the LAN or a specific agent to be updated; (4) a JSON template to specify the LLM's output format. We provide a detailed explanation of this template within the prompt; please refer to Section \ref{section: update step prompt} in the Appendix for more details.

\begin{table}[ht]
\caption{The prompts used in the update steps. Please refer to Section \ref{section: update step prompt} in the Appendix for more details. We omit some punctuation marks in the JSON template.}
\label{table: step prompt}
\resizebox{\textwidth}{!}{%
\begin{tabular}{|c|c|c|c|c|c|}
\hline
 &
  Task for the Step &
  \begin{tabular}[c]{@{}c@{}}Inputs\\ from Previous Steps\end{tabular} &
  \begin{tabular}[c]{@{}c@{}}LAN\\ Description\end{tabular} &
  \begin{tabular}[c]{@{}c@{}}Agent\\ Description\end{tabular} &
  JSON Template \\ \hline
Step 1 &
  \begin{tabular}[c]{@{}c@{}}"Find the gap between the \\ LAN's output and the ground truth"\end{tabular} &
  - &
  Yes &
  No &
  "gap": ... \\ \hline
Step 2 &
  "Find why the gap exists" &
  1. the gap &
  Yes &
  No &
  \begin{tabular}[c]{@{}c@{}}"reason\_type": ...\\ "agent\_name: ...\\ "reason\_content": ...\end{tabular} \\ \hline
Step 3 &
  "Why the agent has a poor performance" &
  \begin{tabular}[c]{@{}c@{}}1. description of \\ the poor performance\end{tabular} &
  No &
  Yes &
  \begin{tabular}[c]{@{}c@{}}"reason\_type": ...\\ "reason\_content": ...\end{tabular} \\ \hline
Step 4 &
  "Calculate the parameter for the strategy" &
  \begin{tabular}[c]{@{}c@{}}1. the agent to be updated\\ (if any)\\ 2. the selected strategy\end{tabular} &
  Yes &
  Yes &
  "parameters": ... \\ \hline
\end{tabular}%
}
\end{table}

\subsection{System Implementation}
We implemented the server back end using Flask and made remote calls to OpenAI's GPT-4-0613. The connection to OpenAI can sometimes be unstable, although this issue is somewhat mitigated after we enabled the "stream" attribute \footnote{\url{https://platform.openai.com/docs/api-reference/completions/create\#stream}}. If the connection breaks down, we continue to retry until OpenAI returns a complete result. We employed React to implement the front end in the browser, and GoJS was utilized for rendering the diagram in Region 1.

\section{Evaluation Study}
\subsection{Experimental Tasks}\label{section: tasks}
Table \ref{table: study tasks} presents the four experimental tasks for the evaluation study. The first was used for the tutorial, and the remaining three were for the formal study. Given that subjective opinions on these tasks could vary among participants, we aimed to reduce variability by establishing well-defined criteria for each task. These criteria were used to evaluate the performance of the LAN. The construction of training samples is beyond the scope of this paper. In this experiment, we constructed 16 training examples for each task, all conformed to the abovementioned criteria. These examples were then split randomly into two subsets: one for training (8 examples) and another for testing (8 examples).

While these tasks are all typical natural language processing tasks, they vary in difficulty for LLMs. LLMs \cite{scao2022bloom, chowdhery2022palm} are trained on multilingual data and excel at translation. However, researchers do not specifically collect sentence compression and couplet generation data to train LLMs. Besides, these tasks are not downstream training objectives for LLMs, making them more challenging.

\begin{table}[ht]
\caption{Tasks in the user study. Please refer to Section \ref{section: tasks} in the Appendix for more details about the tasks.}
\label{table: study tasks}
\resizebox{\textwidth}{!}{%
\begin{tabular}{|c|c|l|}
\hline
\textbf{Task ID} &
  \textbf{Task Description} &
  \multicolumn{1}{c|}{\textbf{Criteria}} \\ \hline
Tutorial &
  Grammatical error detection \cite{liu2021neural} &
  \begin{tabular}[c]{@{}l@{}}1. Identify subject-verb agreement errors\\ 2. Identify missing sentence components\\ 3. Identify redundancy in the sentence\\ 4. Identify ambiguity in the sentence\end{tabular} \\ \hline
1 &
  \begin{tabular}[c]{@{}c@{}}Chinese-English\\ translation \cite{wang2022breaking}\end{tabular} &
  \begin{tabular}[c]{@{}l@{}}1. The result is a grammatically correct sentence.\\ 2. Correctly chooses whether to rhyme or not.\\ 3. Accurately translates words with multiple meanings.\\ 4. Correctly interprets metaphors specific to the Chinese context.\end{tabular} \\ \hline
2 &
  Sentence compression \cite{zi2021som} &
  \begin{tabular}[c]{@{}l@{}}1. The result is a grammatically correct sentence.\\ 2. Remove all the attributives from the sentence.\\ 3. Remove all the adverbials from the sentence.\\ 4. Remove all the complements from the sentence.\end{tabular} \\ \hline
3 &
  \begin{tabular}[c]{@{}c@{}}Chinese couplet\\ generation \cite{chiang2021transcouplet}\end{tabular} &
  \begin{tabular}[c]{@{}l@{}}1. Same character count for both couplets.\\ 2. Matching emotion and related themes.\\ 3. Same parts of speech for corresponding words.\\ 4. No repeating characters between couplets.\end{tabular} \\ \hline
\end{tabular}%
}
\end{table}

\subsection{Participants}
We recruited 12 participants (6 males and 6 females, ages 18-29). All were familiar with the aforementioned tasks, and none were programmers.

\subsection{Baseline}
We use \name{} with its automatic update pipeline disabled as our baseline. Participants can only manually modify the LAN in Regions 1 and 2 (Figure \ref{fig: ui}). They can provide inputs to the LAN and obtain outputs in Region 3. We disable Region 4 in the baseline to prohibit any user interactions.

\subsection{Procedure}
Participants were randomly assigned to two groups to control for individual differences: one group used \name{}, while the other group used the baseline system. We provided participants with a system briefing and explained the experiment's tasks. During the tutorial task, participants received guidance from experimenters and had the opportunity to ask questions.

In the formal study, participants completed three tasks in a randomized order. Training examples were presented randomly, and participants were instructed to modify the LAN until the system's outputs met predefined criteria. The entire experiment lasted approximately 1.5 hours. Afterward, we administered questionnaires to collect subjective feedback.

After conducting experiments with all participants, we fed the testing examples into the user-constructed LANs. during the offline evaluation. The results were then shuffled and independently assessed by two experimenters to determine if they met the predefined criteria. In cases where the opinions diverged, a conclusion was reached through further discussion.

\subsection{Results}
\subsubsection{Terminology and metrics}\ 

\textbf{A LAN modification} is defined as the aggregate of all updates executed between two successive runs of the LAN. This encapsulates the implementation of a comprehensive LAN update plan and may contain multiple operational actions (e.g., creating a new agent and then editing its attributes).

\textbf{User editing distance} (UED): A metric used to estimate the "quantity" of user interactions. We consider a single mouse click or pressing a key (e.g., typing a letter) to have an editing distance of 1. Dragging and text selection are assigned an editing distance of 2, as the user needs to determine both a starting and ending point. Therefore, deleting a text segment has an editing distance of 3, which involves selecting the text first and pressing the "Delete" key.

\textbf{LAN modification distance} (LMD): A metric designed to quantify the differences between two LANs. Let A and B represent two different LANs. If a user manually edits A into B solely through interactions supported by the baseline, the \textit{theoretical minimum} UED required for this transformation is designated as the LMD between A and B. The LMD of a task is defined as the LMD between the initial LAN and the final LAN corresponding to that task.

\textbf{Interaction time}: Defined as the total time minus the LLM execution time. The running of LLM is computationally expensive, and optimizing its efficiency falls outside the scope of this paper. The exclusion of LLM's execution time allows for a more accurate measure of \name{}'s operational efficiency.

\textbf{LAN output scores}: As discussed in Section \ref{section: tasks}, we propose criteria for all tasks. The score of a LAN output is 1 if it fulfills all the criteria, else 0. The score of a LAN is denoted as the average score of its output.

\textbf{Naive LAN}: A LAN with only one agent. The agent relies solely on few-shot learning, meaning that its EM knowledge is exactly the training examples.

\subsection{User behaviors}
\name{} offers a reduced interaction burden compared to the baseline. Table \ref{table: UED} displays the frequency of different editing actions and their average UEDs. In \name{}, the UED to complete a task is 374 (47 per training example), and that for the baseline is 947 (118 per training example, p < 0.001). This suggests that \name{} enables users to develop a satisfactory LAN with 39.5\% of interactions, which is mainly due to two key factors. (1)  \name{} can amplify user actions. The average LMD per task in \name{} is 2424, 6.48 times greater than the corresponding UED. The LMDs for \name{} are very large, mainly because the LLM tended generate very long sentences as knowledge. (2) In the baseline system, the LMD is slightly less than the UED (825 < 945), which indicates that manual edits always introduce errors, necessitating additional corrective actions.

\begin{table}[ht]
\centering
\caption{User actions and their average UED. The UED of \name{} decreases by 60.5\% relative to that of the baseline.}
\label{table: UED}
\tabcolsep=4pt
\begin{tabular}{|c|ccc|ccc|}
\hline
\multirow{2}{*}{Actions} &
  \multicolumn{3}{c|}{Baseline} &
  \multicolumn{3}{c|}{EasyLAN} \\ \cline{2-7} 
 &
  \multicolumn{1}{c|}{Count per task} &
  \multicolumn{1}{c|}{UED per action} &
  UED per task &
  \multicolumn{1}{c|}{Count per task} &
  \multicolumn{1}{c|}{UED per action} &
  UED per task \\ \hline
Modify Step1 &
  \multicolumn{1}{c|}{0} &
  \multicolumn{1}{c|}{0} &
  \multirow{4}{*}{\begin{tabular}[c]{@{}c@{}}0\\ (0\%)\end{tabular}} &
  \multicolumn{1}{c|}{3.33} &
  \multicolumn{1}{c|}{82.4} &
  \multirow{4}{*}{\begin{tabular}[c]{@{}c@{}}301\\ (80.6\%)\end{tabular}} \\ \cline{1-3} \cline{5-6}
Modify Step2 &
  \multicolumn{1}{c|}{0} &
  \multicolumn{1}{c|}{0} &
   &
  \multicolumn{1}{c|}{1.00} &
  \multicolumn{1}{c|}{2.61} &
   \\ \cline{1-3} \cline{5-6}
Modify Step3 &
  \multicolumn{1}{c|}{0} &
  \multicolumn{1}{c|}{0} &
   &
  \multicolumn{1}{c|}{0.94} &
  \multicolumn{1}{c|}{1} &
   \\ \cline{1-3} \cline{5-6}
Modify Step4 &
  \multicolumn{1}{c|}{0} &
  \multicolumn{1}{c|}{0} &
   &
  \multicolumn{1}{c|}{1.39} &
  \multicolumn{1}{c|}{17.0} &
   \\ \hline
New Agent &
  \multicolumn{1}{c|}{1.56} &
  \multicolumn{1}{c|}{1.00} &
  \multirow{3}{*}{\begin{tabular}[c]{@{}c@{}}946\\ (100\%)\end{tabular}} &
  \multicolumn{1}{c|}{0} &
  \multicolumn{1}{c|}{0} &
  \multirow{3}{*}{\begin{tabular}[c]{@{}c@{}}72.3\\ (19.7\%)\end{tabular}} \\ \cline{1-3} \cline{5-6}
Modify Agent &
  \multicolumn{1}{c|}{11.6} &
  \multicolumn{1}{c|}{80.8} &
   &
  \multicolumn{1}{c|}{1.67} &
  \multicolumn{1}{c|}{41.9} &
   \\ \cline{1-3} \cline{5-6}
Modify Edge &
  \multicolumn{1}{c|}{2.11} &
  \multicolumn{1}{c|}{3.16} &
   &
  \multicolumn{1}{c|}{0.33} &
  \multicolumn{1}{c|}{7.00} &
   \\ \hline
\end{tabular}%
\end{table}

\name{} demonstrated a lower frequency of manual modifications, implying that it helped users determine the correct modification plan more quickly. \name{} users conducted 8.22 modifications per task, whereas baseline users conducted 10.83 modifications. The average number of modifications per sample for \name{} is significantly lower than that for the baseline (p=0.048 < 0.05), suggesting that \name{} aided users in making more accurate adjustments to the LAN.

\name{} exhibits higher interaction efficiency. Users completed a task with \name{} in an average interaction time of 605 seconds and saved 39.3\% of the time compared with the baseline (997 seconds, p < 0.001). \name{} exhibits varying degrees of time savings in different tasks. In Task 2 (682 seconds vs. 1046 seconds, p < 0.001) and Task 3 (551 seconds vs. 1325 seconds, p < 0.001), it saved 34.8\% and 58.4\% of time, significantly enhancing interaction efficiency. However, in Task 1, although the average interaction time decreased slightly compared to the baseline (581 seconds vs 620 seconds, 6.3\%), statistical tests did not show significance (p = 0.76). The main reason is that Task 1 is a translation task in which LLM excels, requiring only minor LAN modifications to achieve the desired results. Consequently, the baseline's interaction time is small. In contrast, LLM struggles with Task 2 and Task 3, leading users to make substantial LAN modifications, thus resulting in a significant difference in interaction time between \name{} and the baseline.

\subsection{Performance of the LANs}

LANs constructed using \name{} have a 39.8\% improvement in scores on training samples compared to the baseline (0.583 vs. 0.417, p=0.009 < 0.05). Note that the score of the naive LAN is 0.29 and the tasks are quite difficult for the LLM. The differences mainly exist in Task 1 (0.686 vs. 0.396) and Task 3 (0.563 vs. 0.375). However, the performance of \name{} and the baseline only differs slightly in Task 2 (0.500 vs. 0.479). A possible explanation is that Task 3 is quite suitable for few-shot learning, as the score of the naive LAN in Task 3 is 0.5. Baseline users tended to add the training examples into the LAN as knowledge (we will discuss this in the next paragraph), which explains its performance.


We conducted a thorough comparison of the LANs generated by the two systems and identified the following differences:
\begin{enumerate}
    \item LANs constructed with \name{} exhibit a higher number of agents compared to the baseline system (2.89 vs 2.33, p=0.04 < 0.05), suggesting that \name{} breaks down complex tasks into more detailed components. This is likely because creating a new agent in the baseline system is a cumbersome process for users, as it requires them to manually define various attributes of the agent and edit the values in the GUI.
    \item The knowledge in the LANs constructed by \name{} is more general than that of LANs constructed with the baseline. All of the 121 pieces of knowledge from LANs built with \name{} contain
    both a general description and a concrete example. In contrast, the LAN constructed by the baseline system contains 98 pieces of knowledge, of which 33 (33.7\%) are directly the given training examples (or part of them), without any generalization. Users of the baseline system often found it challenging to generalize knowledge from the training examples, and they tended to add the examples to the knowledge base directly. Although the correctness of the LAN can be ensured during the training process, it is difficult to guarantee the generalization ability.
\end{enumerate}

\subsection{Subjective Feedback}
\begin{table}[ht]
\centering
\caption{Subjective feedback}
\label{table: subjective feedback}
\begin{tabular}{|l|l|l|l|}
\hline
                                                              & Baseline & EasyLAN & P-value \\ \hline
Determining the modification plan for the agent network is easy.       & 3.67 & 5.67 & 0.02  \\ \hline
Applying the modification plan for the agent network is fast  & 4.17     & 6.00    & 0.03    \\ \hline
Interaction with the system is straightforward.               & 4.17     & 6.33    & 0.002   \\ \hline
The system is intelligent, and you receive many hints from the system. & 4.00 & 6.67 & 0.007 \\ \hline
Willing to use this system to construct multi-agent networks. & 4.17     & 6.17    & 0.01    \\ \hline
\end{tabular}%
\end{table}

Table \ref{table: subjective feedback} compares user feedback for two systems. Overall, users have a positive attitude towards \name{}, considering it more intelligent and providing a relaxed and fast interaction experience.

\section{Discussion}
\subsection{The Difference Between Intra-Agent Knowledge and Inter-Agent Structures}
In tutorials of the evaluation study, a frequently asked question is, "When should I add knowledge, and when should I create a new agent?" In fact, the internal structure of a LAN can be understood as encapsulating a form of knowledge that specifies "what actions to take" (hereafter termed W-knowledge). On the other hand, another form of knowledge outlines "how to execute those actions" (hereafter termed H-knowledge). Differentiating between these two categories of knowledge is crucial, as they present distinct characteristics:
\begin{enumerate}
    \item W-knowledge exhibits complex interrelationships, as distinct steps required for accomplishing an intricate task are often interdependent. In the context of LLMs, which solely accept natural language inputs, incorporating multiple pieces of W-knowledge into a single agent necessitates additional natural language descriptions to articulate their interrelationships. Notably, these descriptions serve not as imperatives but as optional constraints. By making the interdependencies among W-knowledge explicit in the structure of the LAN, we achieve a more transparent representation of the hierarchical organization of the subtasks.
    \item W-knowledge is primarily task-oriented, whereas H-knowledge is predominantly input-related. The vast majority of agents are activated during LAN execution, which means that most of the W-knowledge within the LAN structure is fully leveraged, whatever the input is. On the other hand, the relevance of H-knowledge is dependent on specific user input, as illustrated in Table \ref{table: EM property}; for example, the piece of knowledge "levés means lifted in English" is not necessary when the input does not contain "levés". Given the limitations in prompt length and computational capabilities of LLMs, it is impractical to indefinitely incorporate knowledge and expect multitasking proficiency in a single text completion session. To optimize H-knowledge, one can filter relevant pieces based on the input. For managing W-knowledge, a proven strategy, as demonstrated by \name{}, is distributing it among a specialized network of agents, essentially transforming a single agent into a multi-agent network.
\end{enumerate}

However, representing W-knowledge through agents and network architecture imposes a trade-off in computational efficiency. Extending the prompt with an additional sentence generally exerts minimal impact on operational efficiency, whereas invoking an extra LLM computation escalates computational overhead. In our user study, we noted that developers adapt their strategies based on specific scenarios. Initially, they may incorporate W-knowledge into an existing agent; if this fails to enhance the LAN's performance, they adjust the network architecture by introducing new agents or subdividing the existing ones.

\subsection{Capability Boundary of the LAN: the Control Granularity Challenge}\label{section: granularity challenge}
Although users can break down complex tasks into smaller granules to improve the capabilities of LLMs, there are inherent limitations to such granularity. Users could not break down tasks into small granularities that are not practical or meaningful in regular human cognition. A typical example is the couplet generation task. Couplets require that the number of characters in the upper and lower halves match. However, one user found that the LLM's ability to count the number of characters was unstable. This problem might be attributed to the process of encoding and tokenization, which is beyond the user's knowledge. The user could not address this issue within the multi-agent framework and complained, "If it can't even get this simple thing right, I don't know how to fix it." This indicates that optimizing fundamental tasks like counting numbers exceeds the capability of \name{} because those tasks fall below the minimum granularity where humans and LLMs can understand each other.

\section{Limitation \& Future work}
The construction of appropriate training examples falls outside the scope of this paper. Future work could qualitatively and quantitatively investigate the relationship between the features of training examples and the performance of \name{}. Additionally, employing natural language processing techniques for the automated generation or augmentation of training data \cite{dai2023auggpt} represents a promising avenue for future research.

\name{} can only handle acyclic networks, which implies that it cannot accommodate tasks requiring recursion and backtracking. Cycles significantly increase network complexity, as the same agent may be activated multiple times. \name{} faces challenges when confronted with excessively intricate network structures due to the long-distance dependency problem of LLMs. Future work may draw inspiration from existing neural network update strategies (e.g., backpropagation) to handle larger-scale networks.

Currently, \name{} cannot learn from users to improve its performance\cite{pan2022automatically}. Future work can explore how \name{} continuously accumulates knowledge and adapts its pipeline during user interactions. This approach would enable \name{} to autonomously generate a LAN without requiring any user intervention.

\name{} is a collaborative editing tool, not a LAN generation tool. Therefore, it relies on LAN developers rather than on carefully designed strategies to ensure that the constructed LAN is optimal. Future work can focus on improving the LAN's performance by (1) minimizing the number of agents while maintaining performance to improve computational efficiency and (2) adding runtime error monitoring and handling mechanisms to prevent error propagation.

\section{Conclusion}
In this paper, we propose EasyLAN to assist users in developing LLM agent networks for their complex tasks. The most significant feature of EasyLAN is its ability to transform a single agent into an agent network using a small number of training samples. Users monitor and intervene in this process to ensure the accuracy and functionality of EasyLAN. The evaluation study demonstrates that EasyLAN can help users reduce interaction time by 39.3\% while improving the performance of the constructed LANs by 39.8\%. EasyLAN's contributions are two-fold: (1) a novel approach to constructing LLM networks that eliminates the need for users to design network structures and provides assistance during network updates; (2) a novel human-computer collaboration paradigm, "computer execution, user supervision", that reduces the user's burden while ensuring the computational system's correctness. We hope that EasyLAN's preliminary exploration of the LLM agent network can inspire new applications and human-computer collaborative systems in the rapidly evolving era of AI.

\bibliographystyle{unsrt}  
\bibliography{references}

\begin{thebibliography}{10}

\bibitem{khot2022decomposed}
Tushar Khot, Harsh Trivedi, Matthew Finlayson, Yao Fu, Kyle Richardson, Peter Clark, and Ashish Sabharwal.
\newblock Decomposed prompting: A modular approach for solving complex tasks.
\newblock {\em arXiv preprint arXiv:2210.02406}, 2022.

\bibitem{zhou2022least}
Denny Zhou, Nathanael Sch{\"a}rli, Le~Hou, Jason Wei, Nathan Scales, Xuezhi Wang, Dale Schuurmans, Claire Cui, Olivier Bousquet, Quoc Le, et~al.
\newblock Least-to-most prompting enables complex reasoning in large language models.
\newblock {\em arXiv preprint arXiv:2205.10625}, 2022.

\bibitem{dua-etal-2022-successive}
Dheeru Dua, Shivanshu Gupta, Sameer Singh, and Matt Gardner.
\newblock Successive prompting for decomposing complex questions.
\newblock In {\em Proceedings of the 2022 Conference on Empirical Methods in Natural Language Processing}, pages 1251--1265, Abu Dhabi, United Arab Emirates, December 2022. Association for Computational Linguistics.

\bibitem{wu2022promptchainer}
Tongshuang Wu, Ellen Jiang, Aaron Donsbach, Jeff Gray, Alejandra Molina, Michael Terry, and Carrie~J Cai.
\newblock Promptchainer: Chaining large language model prompts through visual programming.
\newblock In {\em CHI Conference on Human Factors in Computing Systems Extended Abstracts}, pages 1--10, 2022.

\bibitem{wu2022ai}
Tongshuang Wu, Michael Terry, and Carrie~Jun Cai.
\newblock Ai chains: Transparent and controllable human-ai interaction by chaining large language model prompts.
\newblock In {\em Proceedings of the 2022 CHI conference on human factors in computing systems}, pages 1--22, 2022.

\bibitem{kochhar1990user}
Sandeep Kochhar and Mark Friedell.
\newblock User control in cooperative computer-aided design.
\newblock In {\em Proceedings of the 3rd annual ACM SIGGRAPH symposium on User interface software and technology}, pages 143--151, 1990.

\bibitem{kochhar1994interaction}
Sandeep Kochhar, Mark Friedell, Joe Marks, Steve Sistare, and Louis Weitzman.
\newblock Interaction paradigms for human-computer cooperation in design.
\newblock In {\em Conference companion on Human factors in computing systems}, pages 187--188, 1994.

\bibitem{liu2023wants}
Michael~Xieyang Liu, Advait Sarkar, Carina Negreanu, Benjamin Zorn, Jack Williams, Neil Toronto, and Andrew~D Gordon.
\newblock “what it wants me to say”: Bridging the abstraction gap between end-user programmers and code-generating large language models.
\newblock In {\em Proceedings of the 2023 CHI Conference on Human Factors in Computing Systems}, pages 1--31, 2023.

\bibitem{chung2022talebrush}
John Joon~Young Chung, Wooseok Kim, Kang~Min Yoo, Hwaran Lee, Eytan Adar, and Minsuk Chang.
\newblock Talebrush: Sketching stories with generative pretrained language models.
\newblock In {\em Proceedings of the 2022 CHI Conference on Human Factors in Computing Systems}, pages 1--19, 2022.

\bibitem{kim2022stylette}
Tae~Soo Kim, DaEun Choi, Yoonseo Choi, and Juho Kim.
\newblock Stylette: Styling the web with natural language.
\newblock In {\em Proceedings of the 2022 CHI Conference on Human Factors in Computing Systems}, pages 1--17, 2022.

\bibitem{dai2021knowledge}
Damai Dai, Li~Dong, Yaru Hao, Zhifang Sui, Baobao Chang, and Furu Wei.
\newblock Knowledge neurons in pretrained transformers.
\newblock {\em arXiv preprint arXiv:2104.08696}, 2021.

\bibitem{meng2022locating}
Kevin Meng, David Bau, Alex Andonian, and Yonatan Belinkov.
\newblock Locating and editing factual associations in gpt.
\newblock {\em Advances in Neural Information Processing Systems}, 35:17359--17372, 2022.

\bibitem{huang2022towards}
Jie Huang and Kevin Chen-Chuan Chang.
\newblock Towards reasoning in large language models: A survey.
\newblock {\em arXiv preprint arXiv:2212.10403}, 2022.

\bibitem{qiao2022reasoning}
Shuofei Qiao, Yixin Ou, Ningyu Zhang, Xiang Chen, Yunzhi Yao, Shumin Deng, Chuanqi Tan, Fei Huang, and Huajun Chen.
\newblock Reasoning with language model prompting: A survey.
\newblock {\em arXiv preprint arXiv:2212.09597}, 2022.

\bibitem{teevan2016productivity}
Jaime Teevan, Shamsi~T Iqbal, Carrie~J Cai, Jeffrey~P Bigham, Michael~S Bernstein, and Elizabeth~M Gerber.
\newblock Productivity decomposed: Getting big things done with little microtasks.
\newblock In {\em Proceedings of the 2016 CHI Conference Extended Abstracts on Human Factors in Computing Systems}, pages 3500--3507, 2016.

\bibitem{bernstein2010soylent}
Michael~S Bernstein, Greg Little, Robert~C Miller, Bj{\"o}rn Hartmann, Mark~S Ackerman, David~R Karger, David Crowell, and Katrina Panovich.
\newblock Soylent: a word processor with a crowd inside.
\newblock In {\em Proceedings of the 23nd annual ACM symposium on User interface software and technology}, pages 313--322, 2010.

\bibitem{zhang2021learning}
Yi~Zhang, Sujay~Kumar Jauhar, Julia Kiseleva, Ryen White, and Dan Roth.
\newblock Learning to decompose and organize complex tasks.
\newblock In {\em Proceedings of the 2021 Conference of the North American Chapter of the Association for Computational Linguistics: Human Language Technologies}, pages 2726--2735, 2021.

\bibitem{wei2022chain}
Jason Wei, Xuezhi Wang, Dale Schuurmans, Maarten Bosma, Fei Xia, Ed~Chi, Quoc~V Le, Denny Zhou, et~al.
\newblock Chain-of-thought prompting elicits reasoning in large language models.
\newblock {\em Advances in Neural Information Processing Systems}, 35:24824--24837, 2022.

\bibitem{brohan2023can}
Anthony Brohan, Yevgen Chebotar, Chelsea Finn, Karol Hausman, Alexander Herzog, Daniel Ho, Julian Ibarz, Alex Irpan, Eric Jang, Ryan Julian, et~al.
\newblock Do as i can, not as i say: Grounding language in robotic affordances.
\newblock In {\em Conference on Robot Learning}, pages 287--318. PMLR, 2023.

\bibitem{zhang2022automatic}
Zhuosheng Zhang, Aston Zhang, Mu~Li, and Alex Smola.
\newblock Automatic chain of thought prompting in large language models.
\newblock {\em arXiv preprint arXiv:2210.03493}, 2022.

\bibitem{kim2017mechanical}
Joy Kim, Sarah Sterman, Allegra Argent~Beal Cohen, and Michael~S Bernstein.
\newblock Mechanical novel: Crowdsourcing complex work through reflection and revision.
\newblock In {\em Proceedings of the 2017 acm conference on computer supported cooperative work and social computing}, pages 233--245, 2017.

\bibitem{zhang2012human}
Haoqi Zhang, Edith Law, Rob Miller, Krzysztof Gajos, David Parkes, and Eric Horvitz.
\newblock Human computation tasks with global constraints.
\newblock In {\em Proceedings of the SIGCHI Conference on Human Factors in Computing Systems}, pages 217--226, 2012.

\bibitem{kaur2018creating}
Harmanpreet Kaur, Alex~C Williams, Anne~Loomis Thompson, Walter~S Lasecki, Shamsi~T Iqbal, and Jaime Teevan.
\newblock Creating better action plans for writing tasks via vocabulary-based planning.
\newblock {\em Proceedings of the ACM on Human-Computer Interaction}, 2(CSCW):1--22, 2018.

\bibitem{zacks2001event}
Jeffrey~M Zacks and Barbara Tversky.
\newblock Event structure in perception and conception.
\newblock {\em Psychological bulletin}, 127(1):3, 2001.

\bibitem{teevan2014selfsourcing}
Jaime Teevan, Daniel~J Liebling, and Walter~S Lasecki.
\newblock Selfsourcing personal tasks.
\newblock In {\em CHI'14 Extended Abstracts on Human Factors in Computing Systems}, pages 2527--2532. 2014.

\bibitem{agapie2016plansourcing}
Elena Agapie, Lucas Colusso, Sean~A Munson, and Gary Hsieh.
\newblock Plansourcing: Generating behavior change plans with friends and crowds.
\newblock In {\em Proceedings of the 19th ACM Conference on Computer-Supported Cooperative Work \& Social Computing}, pages 119--133, 2016.

\bibitem{morris2010people}
Meredith~Ringel Morris, Jaime Teevan, and Katrina Panovich.
\newblock What do people ask their social networks, and why? a survey study of status message q\&a behavior.
\newblock In {\em Proceedings of the SIGCHI conference on Human factors in computing systems}, pages 1739--1748, 2010.

\bibitem{rzeszotarski2014estimating}
Jeffrey~M Rzeszotarski and Meredith~Ringel Morris.
\newblock Estimating the social costs of friendsourcing.
\newblock In {\em Proceedings of the SIGCHI Conference on Human Factors in Computing Systems}, pages 2735--2744, 2014.

\bibitem{anjum2021crowdmot}
Samreen Anjum, Chi Lin, and Danna Gurari.
\newblock Crowdmot: Crowdsourcing strategies for tracking multiple objects in videos.
\newblock {\em Proceedings of the ACM on Human-Computer Interaction}, 4(CSCW3):1--25, 2021.

\bibitem{he2023interaction}
Ziyao He, Yunpeng Song, Shurui Zhou, and Zhongmin Cai.
\newblock Interaction of thoughts: Towards mediating task assignment in human-ai cooperation with a capability-aware shared mental model.
\newblock In {\em Proceedings of the 2023 CHI Conference on Human Factors in Computing Systems}, pages 1--18, 2023.

\bibitem{dang2022prompt}
Hai Dang, Lukas Mecke, Florian Lehmann, Sven Goller, and Daniel Buschek.
\newblock How to prompt? opportunities and challenges of zero- and few-shot learning for human-ai interaction in creative applications of generative models, 2022.

\bibitem{brown2020language}
Tom Brown, Benjamin Mann, Nick Ryder, Melanie Subbiah, Jared~D Kaplan, Prafulla Dhariwal, Arvind Neelakantan, Pranav Shyam, Girish Sastry, Amanda Askell, et~al.
\newblock Language models are few-shot learners.
\newblock {\em Advances in neural information processing systems}, 33:1877--1901, 2020.

\bibitem{reynolds2021prompt}
Laria Reynolds and Kyle McDonell.
\newblock Prompt programming for large language models: Beyond the few-shot paradigm.
\newblock In {\em Extended Abstracts of the 2021 CHI Conference on Human Factors in Computing Systems}, pages 1--7, 2021.

\bibitem{wang2023enabling}
Bryan Wang, Gang Li, and Yang Li.
\newblock Enabling conversational interaction with mobile ui using large language models.
\newblock In {\em Proceedings of the 2023 CHI Conference on Human Factors in Computing Systems}, pages 1--17, 2023.

\bibitem{fiannaca2023programming}
Alexander~J Fiannaca, Chinmay Kulkarni, Carrie~J Cai, and Michael Terry.
\newblock Programming without a programming language: Challenges and opportunities for designing developer tools for prompt programming.
\newblock In {\em Extended Abstracts of the 2023 CHI Conference on Human Factors in Computing Systems}, pages 1--7, 2023.

\bibitem{murali2023improving}
Prasanth Murali, Ian Steenstra, Hye~Sun Yun, Ameneh Shamekhi, and Timothy Bickmore.
\newblock Improving multiparty interactions with a robot using large language models.
\newblock In {\em Extended Abstracts of the 2023 CHI Conference on Human Factors in Computing Systems}, pages 1--8, 2023.

\bibitem{lampinen2022can}
Andrew~K Lampinen, Ishita Dasgupta, Stephanie~CY Chan, Kory Matthewson, Michael~Henry Tessler, Antonia Creswell, James~L McClelland, Jane~X Wang, and Felix Hill.
\newblock Can language models learn from explanations in context?
\newblock {\em arXiv preprint arXiv:2204.02329}, 2022.

\bibitem{su2022selective}
Hongjin Su, Jungo Kasai, Chen~Henry Wu, Weijia Shi, Tianlu Wang, Jiayi Xin, Rui Zhang, Mari Ostendorf, Luke Zettlemoyer, Noah~A Smith, et~al.
\newblock Selective annotation makes language models better few-shot learners.
\newblock {\em arXiv preprint arXiv:2209.01975}, 2022.

\bibitem{deng2022rlprompt}
Mingkai Deng, Jianyu Wang, Cheng-Ping Hsieh, Yihan Wang, Han Guo, Tianmin Shu, Meng Song, Eric~P Xing, and Zhiting Hu.
\newblock Rlprompt: Optimizing discrete text prompts with reinforcement learning.
\newblock {\em arXiv preprint arXiv:2205.12548}, 2022.

\bibitem{zhang2022active}
Yiming Zhang, Shi Feng, and Chenhao Tan.
\newblock Active example selection for in-context learning.
\newblock {\em arXiv preprint arXiv:2211.04486}, 2022.

\bibitem{agrawal2022context}
Sweta Agrawal, Chunting Zhou, Mike Lewis, Luke Zettlemoyer, and Marjan Ghazvininejad.
\newblock In-context examples selection for machine translation.
\newblock {\em arXiv preprint arXiv:2212.02437}, 2022.

\bibitem{liu2021makes}
Jiachang Liu, Dinghan Shen, Yizhe Zhang, Bill Dolan, Lawrence Carin, and Weizhu Chen.
\newblock What makes good in-context examples for gpt-$3 $?
\newblock {\em arXiv preprint arXiv:2101.06804}, 2021.

\bibitem{rubin2021learning}
Ohad Rubin, Jonathan Herzig, and Jonathan Berant.
\newblock Learning to retrieve prompts for in-context learning.
\newblock {\em arXiv preprint arXiv:2112.08633}, 2021.

\bibitem{lu2021fantastically}
Yao Lu, Max Bartolo, Alastair Moore, Sebastian Riedel, and Pontus Stenetorp.
\newblock Fantastically ordered prompts and where to find them: Overcoming few-shot prompt order sensitivity.
\newblock {\em arXiv preprint arXiv:2104.08786}, 2021.

\bibitem{zhao2021calibrate}
Zihao Zhao, Eric Wallace, Shi Feng, Dan Klein, and Sameer Singh.
\newblock Calibrate before use: Improving few-shot performance of language models.
\newblock In {\em International Conference on Machine Learning}, pages 12697--12706. PMLR, 2021.

\bibitem{dong2022survey}
Qingxiu Dong, Lei Li, Damai Dai, Ce~Zheng, Zhiyong Wu, Baobao Chang, Xu~Sun, Jingjing Xu, and Zhifang Sui.
\newblock A survey for in-context learning.
\newblock {\em arXiv preprint arXiv:2301.00234}, 2022.

\bibitem{min2022rethinking}
Sewon Min, Xinxi Lyu, Ari Holtzman, Mikel Artetxe, Mike Lewis, Hannaneh Hajishirzi, and Luke Zettlemoyer.
\newblock Rethinking the role of demonstrations: What makes in-context learning work?
\newblock {\em arXiv preprint arXiv:2202.12837}, 2022.

\bibitem{garg2022can}
Shivam Garg, Dimitris Tsipras, Percy~S Liang, and Gregory Valiant.
\newblock What can transformers learn in-context? a case study of simple function classes.
\newblock {\em Advances in Neural Information Processing Systems}, 35:30583--30598, 2022.

\bibitem{webson2021prompt}
Albert Webson and Ellie Pavlick.
\newblock Do prompt-based models really understand the meaning of their prompts?
\newblock {\em arXiv preprint arXiv:2109.01247}, 2021.

\bibitem{o2021context}
Joe O'Connor and Jacob Andreas.
\newblock What context features can transformer language models use?
\newblock {\em arXiv preprint arXiv:2106.08367}, 2021.

\bibitem{ye2022unreliability}
Xi~Ye and Greg Durrett.
\newblock The unreliability of explanations in few-shot prompting for textual reasoning.
\newblock {\em Advances in neural information processing systems}, 35:30378--30392, 2022.

\bibitem{shi2023large}
Freda Shi, Xinyun Chen, Kanishka Misra, Nathan Scales, David Dohan, Ed~H Chi, Nathanael Sch{\"a}rli, and Denny Zhou.
\newblock Large language models can be easily distracted by irrelevant context.
\newblock In {\em International Conference on Machine Learning}, pages 31210--31227. PMLR, 2023.

\bibitem{turpin2023language}
Miles Turpin, Julian Michael, Ethan Perez, and Samuel~R Bowman.
\newblock Language models don't always say what they think: Unfaithful explanations in chain-of-thought prompting.
\newblock {\em arXiv preprint arXiv:2305.04388}, 2023.

\bibitem{jiang2022promptmaker}
Ellen Jiang, Kristen Olson, Edwin Toh, Alejandra Molina, Aaron Donsbach, Michael Terry, and Carrie~J Cai.
\newblock Promptmaker: Prompt-based prototyping with large language models.
\newblock In {\em CHI Conference on Human Factors in Computing Systems Extended Abstracts}, pages 1--8, 2022.

\bibitem{wilcox2009activenotes}
Lauren Wilcox, Jie Lu, Jennifer Lai, Steven Feiner, and Desmond Jordan.
\newblock Activenotes: computer-assisted creation of patient progress notes.
\newblock In {\em CHI'09 Extended Abstracts on Human Factors in Computing Systems}, pages 3323--3328. 2009.

\bibitem{zhou2017intelligent}
Tianshu Zhou and Jingsong Li.
\newblock An intelligent writing assistant module for narrative clinical records based on named entity recognition and similarity computation.
\newblock 2017.

\bibitem{wang2021brilliant}
Dakuo Wang, Liuping Wang, Zhan Zhang, Ding Wang, Haiyi Zhu, Yvonne Gao, Xiangmin Fan, and Feng Tian.
\newblock “brilliant ai doctor” in rural clinics: Challenges in ai-powered clinical decision support system deployment.
\newblock In {\em Proceedings of the 2021 CHI conference on human factors in computing systems}, pages 1--18, 2021.

\bibitem{sullivan2018tarot}
Anne Sullivan, Mirjam~Palosaari Eladhari, and Michael Cook.
\newblock Tarot-based narrative generation.
\newblock In {\em Proceedings of the 13th International Conference on the Foundations of Digital Games}, pages 1--7, 2018.

\bibitem{zhang2022storybuddy}
Zheng Zhang, Ying Xu, Yanhao Wang, Bingsheng Yao, Daniel Ritchie, Tongshuang Wu, Mo~Yu, Dakuo Wang, and Toby Jia-Jun Li.
\newblock Storybuddy: A human-ai collaborative chatbot for parent-child interactive storytelling with flexible parental involvement.
\newblock In {\em Proceedings of the 2022 CHI Conference on Human Factors in Computing Systems}, pages 1--21, 2022.

\bibitem{davis2016empirically}
Nicholas Davis, Chih-PIn Hsiao, Kunwar Yashraj~Singh, Lisa Li, and Brian Magerko.
\newblock Empirically studying participatory sense-making in abstract drawing with a co-creative cognitive agent.
\newblock In {\em Proceedings of the 21st International Conference on Intelligent User Interfaces}, pages 196--207, 2016.

\bibitem{wang2010idea}
Hao-Chuan Wang, Dan Cosley, and Susan~R Fussell.
\newblock Idea expander: supporting group brainstorming with conversationally triggered visual thinking stimuli.
\newblock In {\em Proceedings of the 2010 ACM conference on Computer supported cooperative work}, pages 103--106, 2010.

\bibitem{guzdial2019friend}
Matthew Guzdial, Nicholas Liao, Jonathan Chen, Shao-Yu Chen, Shukan Shah, Vishwa Shah, Joshua Reno, Gillian Smith, and Mark~O Riedl.
\newblock Friend, collaborator, student, manager: How design of an ai-driven game level editor affects creators.
\newblock In {\em Proceedings of the 2019 CHI conference on human factors in computing systems}, pages 1--13, 2019.

\bibitem{kim2022fitvid}
Jeongyeon Kim, Yubin Choi, Minsuk Kahng, and Juho Kim.
\newblock Fitvid: Responsive and flexible video content adaptation.
\newblock In {\em Proceedings of the 2022 CHI Conference on Human Factors in Computing Systems}, pages 1--16, 2022.

\bibitem{kim2015motif}
Joy Kim, Mira Dontcheva, Wilmot Li, Michael~S Bernstein, and Daniela Steinsapir.
\newblock Motif: Supporting novice creativity through expert patterns.
\newblock In {\em Proceedings of the 33rd Annual ACM Conference on Human Factors in Computing Systems}, pages 1211--1220, 2015.

\bibitem{fan2019collabdraw}
Judith~E Fan, Monica Dinculescu, and David Ha.
\newblock Collabdraw: an environment for collaborative sketching with an artificial agent.
\newblock In {\em Proceedings of the 2019 on Creativity and Cognition}, pages 556--561. 2019.

\bibitem{gatys2017controlling}
Leon~A Gatys, Alexander~S Ecker, Matthias Bethge, Aaron Hertzmann, and Eli Shechtman.
\newblock Controlling perceptual factors in neural style transfer.
\newblock In {\em Proceedings of the IEEE conference on computer vision and pattern recognition}, pages 3985--3993, 2017.

\bibitem{ha2017neural}
David Ha and Douglas Eck.
\newblock A neural representation of sketch drawings.
\newblock {\em arXiv preprint arXiv:1704.03477}, 2017.

\bibitem{karimi2019deep}
Pegah Karimi, Mary~Lou Maher, Nicholas Davis, and Kazjon Grace.
\newblock Deep learning in a computational model for conceptual shifts in a co-creative design system.
\newblock {\em arXiv preprint arXiv:1906.10188}, 2019.

\bibitem{lin2020your}
Yuyu Lin, Jiahao Guo, Yang Chen, Cheng Yao, and Fangtian Ying.
\newblock It is your turn: Collaborative ideation with a co-creative robot through sketch.
\newblock In {\em Proceedings of the 2020 CHI conference on human factors in computing systems}, pages 1--14, 2020.

\bibitem{oh2018lead}
Changhoon Oh, Jungwoo Song, Jinhan Choi, Seonghyeon Kim, Sungwoo Lee, and Bongwon Suh.
\newblock I lead, you help but only with enough details: Understanding user experience of co-creation with artificial intelligence.
\newblock In {\em Proceedings of the 2018 CHI Conference on Human Factors in Computing Systems}, pages 1--13, 2018.

\bibitem{wang2023reprompt}
Yunlong Wang, Shuyuan Shen, and Brian~Y Lim.
\newblock Reprompt: Automatic prompt editing to refine ai-generative art towards precise expressions.
\newblock In {\em Proceedings of the 2023 CHI Conference on Human Factors in Computing Systems}, pages 1--29, 2023.

\bibitem{pan2023human}
Lihang Pan, Chun Yu, Zhe He, and Yuanchun Shi.
\newblock A human-computer collaborative editing tool for conceptual diagrams.
\newblock In {\em Proceedings of the 2023 CHI Conference on Human Factors in Computing Systems}, pages 1--29, 2023.

\bibitem{elgohary2021nl}
Ahmed Elgohary, Christopher Meek, Matthew Richardson, Adam Fourney, Gonzalo Ramos, and Ahmed~Hassan Awadallah.
\newblock Nl-edit: Correcting semantic parse errors through natural language interaction.
\newblock {\em arXiv preprint arXiv:2103.14540}, 2021.

\bibitem{stiennon2020learning}
Nisan Stiennon, Long Ouyang, Jeffrey Wu, Daniel Ziegler, Ryan Lowe, Chelsea Voss, Alec Radford, Dario Amodei, and Paul~F Christiano.
\newblock Learning to summarize with human feedback.
\newblock {\em Advances in Neural Information Processing Systems}, 33:3008--3021, 2020.

\bibitem{ziegler2019fine}
Daniel~M Ziegler, Nisan Stiennon, Jeffrey Wu, Tom~B Brown, Alec Radford, Dario Amodei, Paul Christiano, and Geoffrey Irving.
\newblock Fine-tuning language models from human preferences.
\newblock {\em arXiv preprint arXiv:1909.08593}, 2019.

\bibitem{mosqueira2023human}
Eduardo Mosqueira-Rey, Elena Hern{\'a}ndez-Pereira, David Alonso-R{\'\i}os, Jos{\'e} Bobes-Bascar{\'a}n, and {\'A}ngel Fern{\'a}ndez-Leal.
\newblock Human-in-the-loop machine learning: A state of the art.
\newblock {\em Artificial Intelligence Review}, 56(4):3005--3054, 2023.

\bibitem{wu2022survey}
Xingjiao Wu, Luwei Xiao, Yixuan Sun, Junhang Zhang, Tianlong Ma, and Liang He.
\newblock A survey of human-in-the-loop for machine learning.
\newblock {\em Future Gener. Comput. Syst.}, 135(C):364–381, oct 2022.

\bibitem{scao2022bloom}
Teven~Le Scao, Angela Fan, Christopher Akiki, Ellie Pavlick, Suzana Ili{\'c}, Daniel Hesslow, Roman Castagn{\'e}, Alexandra~Sasha Luccioni, Fran{\c{c}}ois Yvon, Matthias Gall{\'e}, et~al.
\newblock Bloom: A 176b-parameter open-access multilingual language model.
\newblock {\em arXiv preprint arXiv:2211.05100}, 2022.

\bibitem{chowdhery2022palm}
Aakanksha Chowdhery, Sharan Narang, Jacob Devlin, Maarten Bosma, Gaurav Mishra, Adam Roberts, Paul Barham, Hyung~Won Chung, Charles Sutton, Sebastian Gehrmann, et~al.
\newblock Palm: Scaling language modeling with pathways.
\newblock {\em arXiv preprint arXiv:2204.02311}, 2022.

\bibitem{liu2021neural}
Zhenghao Liu, Xiaoyuan Yi, Maosong Sun, Liner Yang, and Tat-Seng Chua.
\newblock Neural quality estimation with multiple hypotheses for grammatical error correction.
\newblock {\em arXiv preprint arXiv:2105.04443}, 2021.

\bibitem{wang2022breaking}
Zhijun Wang, Xuebo Liu, and Min Zhang.
\newblock Breaking the representation bottleneck of chinese characters: Neural machine translation with stroke sequence modeling.
\newblock {\em arXiv preprint arXiv:2211.12781}, 2022.

\bibitem{zi2021som}
Kangli Zi, Shi Wang, Yu~Liu, Jicun Li, Yanan Cao, and Cungen Cao.
\newblock Som-ncscm: An efficient neural chinese sentence compression model enhanced with self-organizing map.
\newblock In {\em Proceedings of the 2021 Conference on Empirical Methods in Natural Language Processing}, pages 403--415, 2021.

\bibitem{chiang2021transcouplet}
Kuan-Yu Chiang, Shihao Lin, Joe Chen, Qian Yin, and Qizhen Jin.
\newblock Transcouplet: Transformer based chinese couplet generation.
\newblock {\em arXiv preprint arXiv:2112.01707}, 2021.

\bibitem{dai2023auggpt}
Haixing Dai, Zhengliang Liu, Wenxiong Liao, Xiaoke Huang, Yihan Cao, Zihao Wu, Lin Zhao, Shaochen Xu, Wei Liu, Ninghao Liu, et~al.
\newblock Auggpt: Leveraging chatgpt for text data augmentation.
\newblock {\em arXiv preprint arXiv:2302.13007}, 2023.

\bibitem{pan2022automatically}
Lihang Pan, Chun Yu, JiaHui Li, Tian Huang, Xiaojun Bi, and Yuanchun Shi.
\newblock Automatically generating and improving voice command interface from operation sequences on smartphones.
\newblock In {\em Proceedings of the 2022 CHI Conference on Human Factors in Computing Systems}, pages 1--21, 2022.

\end{thebibliography}

\end{document}